%% file: demo.tex
\documentclass[a4paper,fleqn]{cas-dc}

\usepackage[numbers]{natbib}
\usepackage{subfigure}

\usepackage{lineno,hyperref}
\usepackage{graphicx,floatrow}
\usepackage{subfigure}
\usepackage{amsfonts}
\usepackage{amssymb}
\usepackage{amsmath}
\usepackage{algorithm}
\usepackage{algorithmic}
\usepackage{makecell,rotating,multirow,diagbox}
\usepackage{epstopdf}
\usepackage{color}
\usepackage{textcomp}
\usepackage{bm}
\usepackage{threeparttable}
\usepackage{booktabs}
\usepackage{graphicx}
\usepackage{longtable}
\usepackage{booktabs}
\usepackage{supertabular}
\usepackage{pdfpages}
\usepackage{calc}
\usepackage{url}
\usepackage{tikz}
\usepackage{lettrine}
\begin{document}
\let\WriteBookmarks\relax
\def\floatpagepagefraction{1}
\def\textpagefraction{.001}
\shorttitle{When Blockchain Meets Smart Grids: A Comprehensive Survey}
\shortauthors{Yihao Guo et~al.}

\title [mode = title]{When Blockchain Meets Smart Grids: A Comprehensive Survey}

\author[1]{Yihao Guo}
\ead{yhguo@mail.sdu.edu.cn}

\author[1]{Zhiguo Wan}
\cormark[1]
\ead{wanzhiguo@sdu.edu.cn}

\author[1]{Xiuzhen Cheng}[orcid=0000-0001-5912-4647]
\ead{xzcheng@sdu.edu.cn}

\address[1]{School Computer Science and Technology, Shandong University, Qingdao, R.P.China}

\cortext[1]{Corresponding author}



\begin{abstract}
	Recent years have witnessed an increasing interest in the blockchain technology, and
	many blockchain-based applications have been developed to take advantage of its
	decentralization, transparency, fault tolerance, and strong security. In the field of
	smart grids, a plethora of proposals
	have emerged to utilize blockchain for augmenting intelligent energy management, energy trading, security and privacy protection,  
	microgrid management, and energy vehicles. Compared with
	traditional centralized approaches, blockchain-based solutions are able to exploit
	the advantages of blockchain to realize better functionality in smart grids.
	However, the blockchain technology itself has its disadvantages in low 
	processing throughput
	and weak privacy protection. Therefore, it is of paramount importance to study
	how to integrate blockchain with smart grids in a more effective way so that the advantages
	of blockchain can be maximized and its disadvantages can be avoided.
	
	This article surveys the state-of-the-art solutions aiming to integrate
	the emergent blockchain technology with smart grids. The goal of this survey is to
	discuss the necessity of applying blockchain in different components of smart grids,
	identify the challenges encountered by current solutions,
	and highlight the frameworks and techniques used to integrate blockchain with smart grids.
	We also present thorough comparison studies among blockchain-based solutions for smart grids
	from different perspectives, with the aim to provide insights on integrating blockchain with smart grids for different smart grid management tasks.
	Finally, we list the current projects and initiatives demonstrating the current effort from the practice side. Additionally, we draw attention to open problems that have not yet been tackled by
	existing solutions, and point out possible future research directions.
\end{abstract}



\maketitle
\input{2Introduction}
\input{3Blockchain}

\input{4management}
\input{5trading}
\input{6security-and-privacy}
\input{7microgrid}
\input{8v2g}

\input{9resource}
\input{10summary}
\input{11conclusion}

\printcredits

\bibliographystyle{cas-model2-names}
\bibliography{demo}


\end{document}

%% file: 2Introduction.tex
\section{Introduction}\label{intro}
The groundbreaking blockchain technology has gained tremendous attention from governments and
companies to research institutions all over the world in recent years.
This is due to the outstanding advantages of blockchain in
decentralization, transparency, immutability, fault tolerance, and strong security.
Essentially, blockchain is a decentralized consensus ledger managed by multiple
maintainers (called miners or validators), without relying on a centralized server.
For instance, the underlying blockchain of Bitcoin or Ethereum is maintained
by thousands of miners using a specific consensus mechanism (e.g. proof-of-work or proof-of-stake).
Each transaction sent to the blockchain is verified and executed by all the validators,
and it cannot be changed anymore after it is confirmed on the blockchain.
Besides transactions, blockchains like Ethereum and Hyperledger fabric
also support the so-called smart contracts, which are self-executable programs running over blockchains.
Consequently, blockchain functions as a highly secure and trustworthy ``consensus computer"
that can serve as a trusted intermediary. 

After the surprising success in the cryptocurrency field, the blockchain technology has been rapidly
introduced to many other fields ranging from financing, banking, insurance, manufacturing,
supply chain management, provenance, healthcare, Internet-of-Things to cloud computing.
More than 70 international banks have joined R3 to develop the open blockchain platform Corda,
aiming to lower cost of financing and banking. The Internet giant Facebook has announced its own cryptocurrency Libra, which has drawn the attention of the whole world. IBM, Amazon, Samsung, and other companies have also shown their strong
interests in blockchain, e.g. IBM's supply chain and Walmart's blockchain. In addition, a number of countries have already started to apply blockchain
to improve transparency in government management. In conclusion, blockchain 
has been employed to empower various applications in many fields. In this article, we focus on the application of blockchain in smart grids.

\subsection{Why Smart Grids Need Blockchain}
A smart grid is a critical infrastructure that can be significantly improved with the blockchain
technology. Current smart grids incorporate communication and control technologies into
the power grids, such that significant improvements on energy efficiency and grid safety can be achieved. To this end, numerous smart devices are distributed throughout a smart grid to efficiently manage power generation, transmission, distribution and consumption.
It is critical to effectively manage these smart devices and maintain security and
stability of the smart grid. However, traditional centralized approaches to manage smart grids
face severe challenges in various aspects, ranging from energy management, electricity trading, security \& privacy, microgrid management, to electric vehicle management, which are further detailed in the following descriptions.

\begin{itemize}
  \item \textbf{Energy Management.} The centralized grid management mode leads to the single point of failure and severe security issues.
  A potential attacker would launch various attacks specifically targeting at a control center to maximize damage.
  For instance, denial-of-service attacks against the control center may result in serious damages to the whole power grid.
  Moreover, the centralized grid management mode cannot make the best of the distributed renewable energy sources, energy storage units, and electric vehicles to achieve efficient energy management.
  It is challenging for the control center to make good and timely scheduling decisions for the highly dynamic and distributed renewable energy sources and storage units.
  At the same time, with the increase number of prosumers, the burden and cost of centralized energy management have been increased, which makes it impossible to achieve effective demand response in many cases.
  Especially for demand-side management, some consumers are more inclined to purchase electricity from prosumers according to their own wishes, rendering energy management even more challenging.

  \item \textbf{Electricity Trading.} The widespread use of distributed energy indicates that the transformation of the centralized smart grid transaction model to a decentralized one is inevitable.
  The centralized mode cannot efficiently deal with energy trading among distributed energy providers, consumers, and prosumers throughout a smart grid.
  Moreover, the price of centralized grid energy transactions cannot meet the expectations of the market, and it is easy to cause unreasonable monopoly of electricity prices.
Therefore, it is important to improve energy trading efficiency and reliability of the smart grid.

  \item \textbf{Security \& Privacy.} There exist growing security concerns on centralized smart grids.
  External malicious attacks, third-party reliance, and privacy leakage have caused huge economic losses of the power grids.
  The traditional centralized power grid has limited solutions.
  Most situations rely on human supervision. However, it is a challenging task to manage the large amount of smart devices for status monitoring, grid control and metering in a centralized manner.
  On the other hand, numerous smart sensors generate huge amount of data, which is hard to be stored and processed by a centralized server.
  Moreover, it is also inefficient for the control center to maintain security for so many smart devices in case of malicious intrusions.

  \item \textbf{Microgrid Management.} As a specific application scenario of the power grid, the microgrid has the characteristics of regional management, solving the problems caused by long-distance energy transmission in a large power grid.
  However, as a sub-system of the large power grid, it still has huge hidden dangers in energy distribution, security and privacy protection, and load balancing.

  \item \textbf{Electric Vehicle Management.} As another application scenario of the power grid, the combination of energy vehicles and the power grid adds a great flexibility to long-distance power transmission.
  Being an independent power storage unit, each vehicle has certain limitations on its computing power and is extremely vulnerable to external attacks.
  Especially in the process of charging/discharging and power transmission, secure payment and demand response are greatly challenged.
  Moreover, in V2G (vehicle-to-grid), because of the need to exchange road conditions, vehicles have higher requirements for real-time performance.
\end{itemize}

\begin{figure}[htbp]
	\includegraphics[width=0.7\linewidth]{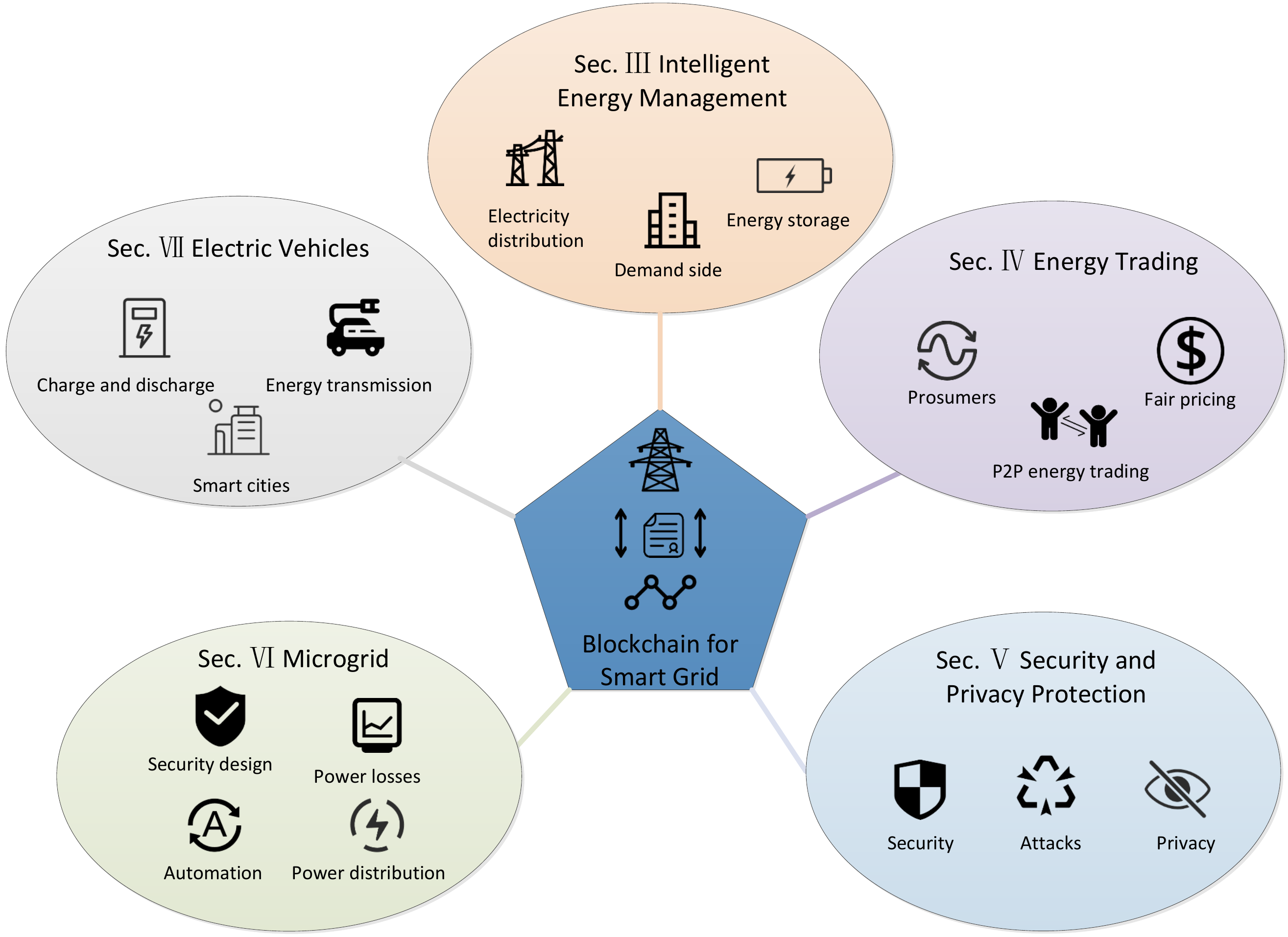}
	\caption{Roadmap of blockchain-based smart grids.}\label{fig:structure}	
\end{figure}

This survey dedicates to the application of the blockchain technology in smart grids for various purposes.
We divide our study into 5 broad areas: energy management, energy trading, security and privacy protection,
microgrid management, and electric vehicles. This corresponds to the five challenges mentioned above. The road map is shown in Fig. \ref{fig:structure}. For each area, we discuss issues in conventional approaches and highlight existing problems and issues therein, describe
the blockchain-based solutions that utilize the blockchain technology to address the corresponding issues, emphasize the applicability of blockchain in these blockchain-based solutions,
and discuss the lessons learned from these studies. We also summarize existing projects and initiatives that integrate blockchain with smart grids from a practical side. Finally we analyze the remaining research challenges and put forward a number of potential research problems.

\subsection{Related works}

A number of surveys have been conducted on related topics recently.
Meisel et al.~\cite{8217069} focus on research works that apply blockchain to microgrids. They study 14 cases of integrating microgrids and blockchain, and analyze the advantages in each case.
Xie et al.~\cite{xie2019survey} investigate the research works on applying blockchain to smart cities, treating a smart grid as a part of a smart city. This survey analyzes about 20 related papers on combination of blockchain and smart grids.
Andoni et al.~\cite{andoni2019blockchain} focus on the development of the energy sector, particularly the emerging projects in the energy sector. They examine 140 emerging blockchain energy enterprise projects, while only summarizing a small number of research papers.
Ref.~\cite{8730307} studies 40 papers on combining blockchain and smart grids, and discusses several important aspects of smart grids, including energy trading, security, electric vehicles, and microgrids. The authors also describe 3 new frameworks and 10 practical projects that integrate blockchain with smart grids.
Alladi et al.~\cite{alladi2019blockchain} focus on the integration of blockchain and smart grids too. This survey considers energy transaction, energy trading based vehicles, security \& privacy, and business cases, provides a number of valuable views, and puts forward the corresponding challenges and prospects.

More recently, Sengupta et al.~\cite{sengupta2020comprehensive} summarize many papers on blockchain and IoT/Industrial IoT, among which 20 of them are on  smart grid security. Both  \cite{ajomand2020review} and \cite{mollah2020blockchain} analyze the related works
from the aspects of energy trading, security, microgrids, and electric vehicles, while ignorning energy management. Ajomand et al.~\cite{ajomand2020review} provide a review on smart grids but they only survey about 20 papers. Ref.~\cite{mollah2020blockchain} offers a survey on about 40 papers and the related projects.

\subsection{Uniqueness of This Survey Article}

The survey articles mentioned in the previous subsection have made outstanding contributions to the understanding of the state-of-the-art in applying the blockchain technology to smart grids. Although smart grid is not the focused research subject in some articles \cite{xie2019survey, andoni2019blockchain, sengupta2020comprehensive}, the results of their collation still have high reference values. However, the shortcomings are also obvious. Particularly, there lacks enough articles to comprehensively summarize the unique characteristics of integrating blockchain with smart grids. Furthermore, smart grid management, which is particularly important, is only slightly mentioned in \cite{andoni2019blockchain}. In addition, some articles only provide a certain concept expression, lacking the description and summary of specific programs \cite{alladi2019blockchain, ajomand2020review}. Finally, the rapid development of blockchain and smart grids requires a more trendy, comprehensive, specific, and innovative article that can summarize the major recent academic achievements.

These factors motivate us to write this article. We intend to provide a comprehensive survey covering as many research works as possible, including the latest papers accepted or published in 2021. We divide the subject of this survey into 5 areas, involving more than 150 specific papers. For each area, we first outline the current challenges, then survey blockchain-based solutions,  and finally provide an analytical summary and comparison study. Particularly, we make a lot of effort to outline the main idea of each paper,  exemplify representative solutions and key technical details in the forms of figures and tables. One can see that the scope of research and the number of papers surveyed by this article are both unprecedented. To further improve the completeness and practicality of this survey, we sort out related sources, including important field experiments and current projects. Moreover, our survey covers the management of smart grids that have been overlooked by most related surveys, and divides it into supply side energy distribution management and demand side management. One can see that this survey article provides a clear logic for readers to comprehensively understand the subject areas, from headlines to subtitles then to the theme of each paragraph.

\subsection{Organization}\label{Organization}

This article is organized as follows. In the next section, we provide a brief overview on the blockchain and smart contract technologies,
emphasizing the opportunities and advantages of the combination of blockchain and smart grids.
In Section \ref{Management}, we present the blockchain-based schemes on energy management, covering energy distribution and demand side management.
Section \ref{trading} discusses the blockchain-based energy trading schemes for smart grids, emphasizing how the pricing of a trading process is ensured.
Then we describe the efforts on applying the blockchain technology to protect security and privacy in Section~\ref{security}.
Section \ref{Microgrid} summarizes the blockchain-based approaches to microgrid management, demonstrating how safety, automation, and power distribution in microgrids are realized.
Section \ref{Vehicle} presents the blockchain-based solutions on management of electric vehicles, including vehicle charges/discharges and energy transmission.
In Section \ref{resource}, we collect the resources on projects and initiatives that integrate blockchain and power grids, focusing on related practical applications and public experiments.
Finally, we discuss open research issues to stimulate further research.

%% file: 3Blockchain.tex
\section{Background}

\subsection{Blockchain and Smart Contract}
The blockchain technology is originated from Bitcoin proposed by Satoshi Nakamoto in 2008. With the deepening of research, Ethereum, Tendermint, IOTA, ZCash, and other blockchain platforms have been gradually developed. They have their own characteristics in terms of delay and functionality, but their essence is a distributed database. In a blockchain system, miners play an important role, both as participants and as maintainers of the blockchain, upholding the system through actively accounting and packaging verified transactions to obtain certain rewards.
The emergence of consensus algorithms improves the stability of blockchain systems and solves the problem of data asynchrony. More importantly, blockchain systems support various consensus algorithms such as PoW, PoS, and PBFT, which possess their own advantages and are suitable for different scenarios.

Blockchain systems can be classified as public blockchains, consortium blockchains, and private blockchains. These three categories differ in the restrictions on the members who can participate. Public blockchains are open to all the people and thus everyone can participate; consortium blockchains only allow alliance members to participate; while private blockchains can only be maintained by a person or an organization. The specific differences are shown in Table \ref{Comparison}. Due to the unique design concepts, blockchain systems have many characteristics such as decentralization, transparency, immutability, and trustworthiness.

The idea of smart contracts is originally proposed by Nick Szabo. A smart contract is essentially a computer protocol to guarantee trusted transactions without the intervention of third parties. The advantages of smart contracts include reducing manpower, lowering costs, and realizing the automation of the system operation. Smart contracts exist in digital forms, which allow them to integrate perfectly with computers. They are generally formulated and executed together with users to better reflect their wishes. So far, not all blockchain platforms can support smart contracts. The most famous combination of smart contracts and blockchain is the implementation of Ethereum. There also exist many blockchain platforms that do not support smart contracts, including Bitcoin, Factom, etc.

\begin{table*}[htbp]
	\begin{threeparttable} \footnotesize
		\caption{Comparison Of Public, Private And Consortium Blockchain}
		\vspace{-0.3cm}
		\label{Comparison}
		\begin{tabular}{c c c c c c c}
			\toprule
			Blockchain   & Participants  &   Access control & Degree of decentralization & Security level & Transaction speed & Transaction cost  \\
			
			\specialrule{0em}{1pt}{1pt}
			\midrule
			Public blockchain &  All users   &  No  & Completely   & High & Slow & High \\
			\specialrule{0em}{1pt}{1pt}
			Private blockchain & Partial users   & Yes  & Partially & Low & Fast & Low \\
			\specialrule{0em}{1pt}{1pt}
			Consortium  blockchain & Partial users  & Yes & Partially & Medium & Fast &  Medium \\
			\bottomrule
		\end{tabular}
	\end{threeparttable} \normalsize
\end{table*}

\subsection{Integrating Blockchain with Smart Grids: Opportunities and Advantages}

\begin{figure*}[ht]
	\includegraphics[width=0.9\linewidth]{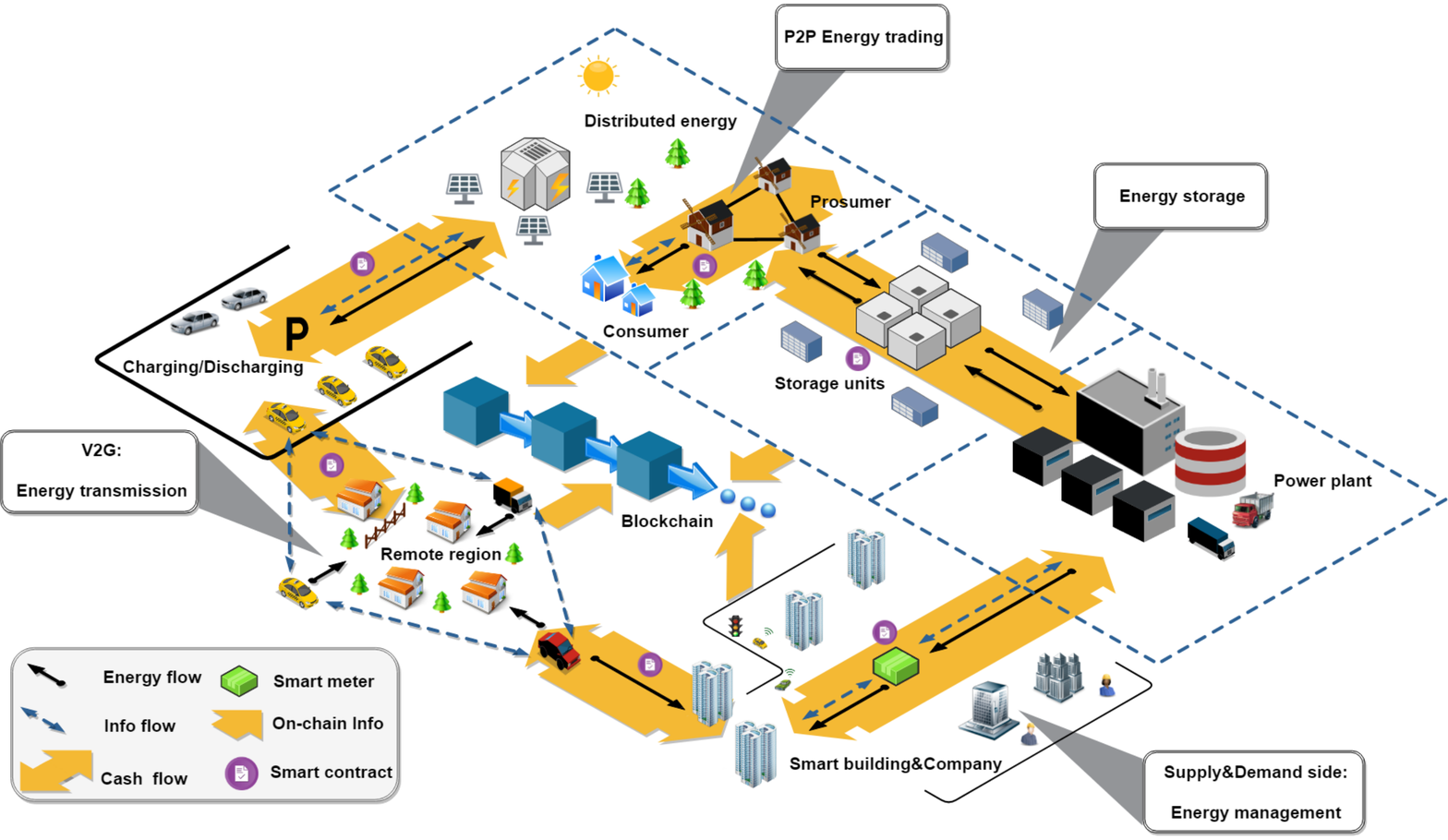}
	\caption{Application scenarios of blockchain and smart grids.}\label{WechatIMG85}
	
\end{figure*}

Smart grid is an intelligent transformation of traditional power grid. It solves a series of problems existing in traditional power grids such as poor elasticity, information islands, and single information flow. In a smart grid, the physical structure and the division of labor are clear. The information of the entire smart grid can be obtained by specialized agencies to achieve optimal distribution of energy. However, a smart grid still has a lot of problems and face a number of challenges (shown in Fig. \ref{intro}). This provides opportunities for blockchain to enter the smart grid field. More specifically,  blockchain, as a new type of distributed structure, can be integrated with a smart grid (as illustrated in Fig. \ref{WechatIMG85}) and has a positive impact on the smart grid.
\begin{itemize}

\item  Blockchain can transform and upgrade centralized grid management to distributed intelligent management. Firstly, a blockchain can realize effective management on the supply side of a grid. For a power grid with a large number of equipment and distributed energy units, the blockchain can provide decentralized supervision, accurate demand response, supply-demand balance, and optimized distribution of electrical energy. This enables the smart grid to cater to the growing trend of prosumers. Secondly, a blockchain can provide more democratic management for the demand side. Users can join the blockchain as nodes for energy data inspection, supervision, and even formulating corresponding rules according to their wishes through smart contracts. Finally, the decentralized nature of a blockchain can transform a smart grid from centralized storage management to distributed multi-point management.  
The smart grid enabled by the blockchain technology can realize effective scheduling of electricity storage to minimize energy loss.

\item  The financial nature of blockchain makes it an ideal platform for electricity trading in smart grids. A blockchain enabled smart grid can realize optimized data flow, energy flow and cash flow in the process of energy trading.
On one hand, energy trading in a traditional grid is usually centralized and thus is easy to cause monopoly. The decentralized nature of blockchain can allow prosumers to join the grid and trade electricity in a P2P manner. The immutability of blockchain can record the process of energy transactions and the interactions of related data. 
Meanwhile, the transparency of blockchain allows users to verify grid data, which makes grid energy transactions and data more open and reliable.
On the other hand, blockchain can provide a secure cash flow for energy trading. Cryptocurrency has proven its security, credibility and convenience in payment processing.
The incentive mechanism of blockchain and smart contracts can realize dynamic pricing and flexible auctions between prosumers. This makes electricity trading in a smart grid more flexible and convenient.

\item Blockchain can greatly improve the security and privacy of the power grid for its decentralization and fault tolerance. A blockchain itself is highly tolerant to faults, which can enhance the underlying security of the power grid. Specifically, the consensus algorithm can effectively enhance the robustness of the system and resist malicious attacks. It is worth mentioning that the blockchain does not rely on third parties, which effectively prevents leaking private information to third parties. Considering the large number of users, the anonymity of blockchain can protect identity information.

\item Blockchain can also extend its advantages to microgrids. Due to the regional characteristics of a microgrid, strict access control mechanisms are required. Blockchains have their own identity management mechanisms, which can be adapted to suite the needs of microgrids. Moreover, a microgrid powered by the blockchain technology can facilitate electricity transmission within a short distance, thereby significantly reducing electricity loss. The combination of smart contracts and microgrids can also make the automated management of microgrids more rational and transparent.

\item Blockchain is well-suited for managing the V2G network in smart grids. The financial nature and security mechanisms of blockchain can ensure information security during the charging and discharging process. Each vehicle is responsible for energy storage and transmission, and can be used as a blockchain node for mining and reaching consensus. Lightweight consensus algorithms require low computing power and are suitable for energy vehicles.

\end{itemize}

%% file: 4management.tex
\section{Blockchain for Intelligent Energy Management}\label{Management}

\subsection{Current Issues in Intelligent Energy Management}


Energy management is the general term for scientific planning, organization, inspection, control and supervision of the entire process of energy production, distribution, conversion and consumption. It can be roughly divided into supplier energy management and demand side energy management.
In order to achieve intelligent smart grid management, a number of schemes have been proposed to analyze consumption data for building models. However, existing methods have poor scalability and require expensive computation for real-time and dynamic smart grids. In addition, traditional energy management has problems with centralized storage, and hence is prone to single-point failures, which is easy to cause excessive  waste of energy resources.


It turns out that solutions based on blockchain can implement effective management of electrical energy. They not only improve security of smart grids, reduce energy consumption and manpower, but also achieve higher scalability and decentralized design. These solutions can be further divided into two groups: effective electricity distribution and demand side management. For these two specific groups, we elaborate on the problems and technical characteristics of each solution. Finally, we conduct a comparative analysis on these solutions.

\subsection{Effective Electricity Distribution}


The goal of energy distribution is to meet electricity demand, so as to achieve reasonable energy management.
Tsolakis et al.~\cite {TsolakisA} establish a secure demand response platform based on OpenADR and blockchain. In this platform, OpenADR can make the equipment respond to peaks and troughs of electricity consumption, thereby reducing power consumption and realizing a reasonable energy distribution. In order to enhance robustness and realize supervision of the system, an information collection center is established in this system.
Like \cite {TsolakisA}, Van et al.~\cite{van2020cooperative} design an energy management platform for smart buildings, aiming to meet the user's demand for renewable energy. At the same time, the authors conduct an experimental test based on Ethereum and employ smart contracts for automatic management and monitoring. The results show that this scheme is suitable in a distributed environment with high power consumption. Edmonds et al.~ \cite{edmonds2020blockchain} propose an energy management platform for smart homes to address demand response. They make use of  smart contracts and propose a privacy protection mechanism based on a private blockchain.


In addition to meeting demand response, effective energy distribution can optimize the entire energy system.
In \cite{alskaif2019decentralized}, AlSkaif et al. apply a distributed optimal power flow model to manage the local distribution network based on a private blockchain.
On this basis, the authors propose an optimization algorithm based on ADMM (alternating
direction method of multipliers), achieving more reasonable and efficient energy distribution. Similarly, Yang et.al~\cite{yang2021blockchain} use ADMM to achieve cost minimization modeling, which is especially suitable for flexible appliance scheduling.

Marius et al.~\cite{BusinessSmart2020} utilize reasonable energy distribution to achieve load optimization in virtual power plants. Moreover, research works~\cite{nakayama2019transactive, yahaya2020blockchain} aim to realize optimal management of energy costs. Ref.~\cite{nakayama2019transactive} uses smart contracts to implement economic scheduling among the prosumers and protect the interests of each prosumer. Yahaya et al.~\cite{yahaya2020blockchain} propose a model based on smart contracts and the demurrage technology to optimize consumers' energy consumption as well as achieve a reasonable distribution of energy.


Blockchain also shows a great potential in managing renewable resources.
Ref. \cite{LazaroiuBlockchain} uses smart contracts to achieve automation and transparent design. Lazaroiu et al. propose a price prediction algorithm to improve the accuracy of energy allocation. Experiments show that this scheme is particularly suitable for smart cities and prosumers. In addition, the main purpose of the scheme \cite{KimSecurity} is to ensure the security and privacy of renewable energy in the energy management process. Kim et al. summarize the potential security risks when renewable energy sources join the traditional power grid distribution management, and propose to use blockchain to improve the management of the smart grid.
Skowronski \cite{SkowronskiOn} proposes an energy management solution for green-field energy and realizes a smart grid based on the GRIDNET protocol. Smart devices in the smart grid are divided into light nodes and full nodes, and the maintenance of the entire blockchain is managed by the full nodes. Each node has a certificate when registering, and all full nodes run PoW consensus algorithm to manage the blockchain. The blockchain is utilized to prevent node misbehaviors and motivate benign behaviors.

Shao et al.~\cite {shao2019grid} propose a DER (Distributed Energy Resources) grid-connected VPP (virtual power plant) model. This model combines with the incentive measures of DER and blockchain to realize a reasonable energy distribution. In~\cite{wang2018blockchainassisted},
a secure crowdsourcing energy platform is proposed to adjust energy distribution. In this scheme, crowdsourcing sources come from two types: regular submissions of crowdsourcing tasks and real-time submissions. In order to motivate users, Wang et al.~\cite{wang2018blockchainassisted} design a monetary incentive mechanism, which is based on market demand and game theory.
Furthermore, Utz et al.~\cite{utz2018blockchain} implement an energy management platform based on smart contracts to meet the distributed problems caused by prosumers. This platform also implements related incentive mechanisms to facilitate rational use of distributed energy.

\begin{figure}[htbp]
	\includegraphics[width=0.95\linewidth]{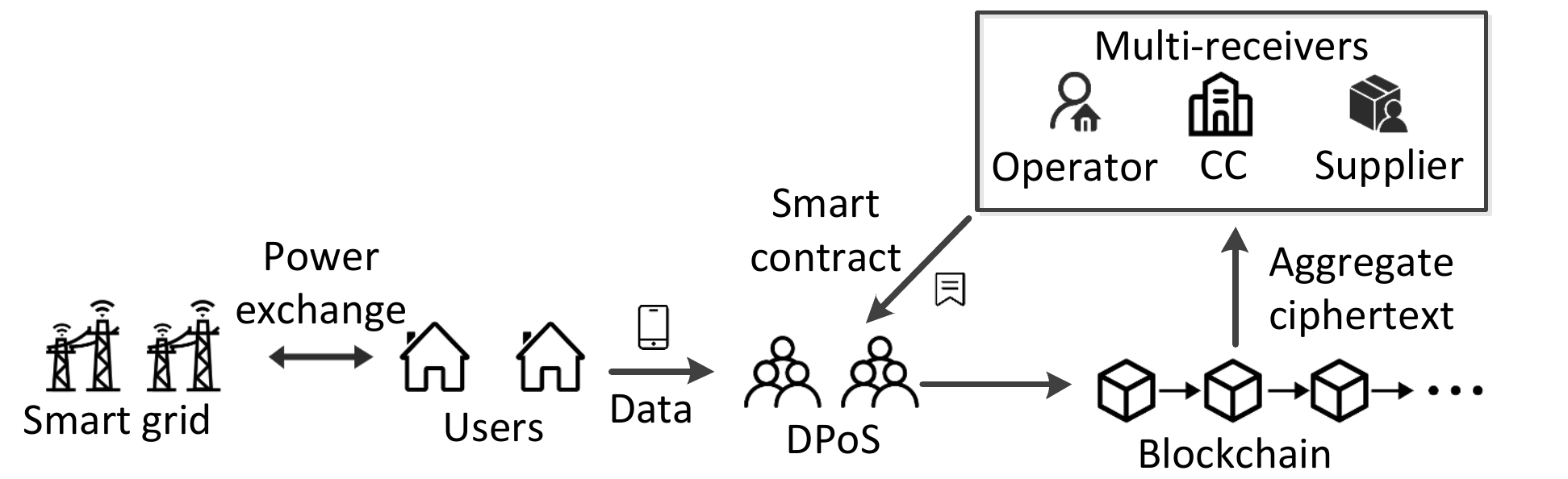}
	\caption{ The workflow description of CBSG \cite{FanConsortium}.}
	\label{fig:FanConsortium}
\end{figure}

CBSG~\cite{FanConsortium} is designed to monitor and manage the data generated during energy distributions. It combines the data regulation, signcryption, and aggregation algorithms,  ensuring the safety and integrity of energy data (shown in Figure \ref{fig:FanConsortium}).
Suciu et al.~\cite{suciu2019blockchain} propose SealedGrid to provide effective distributed energy management. On this basis, SealedGrid provides real-time monitoring functions to create a secure environment for participants.

\subsection{Demand Side Management}


Blockchain based demand side management must first solve the problems of price and cost optimization to achieve economic savings.
Afzal et al.~\cite{afzal2020blockchain} apply smart contracts to realize automated billing functions in the community, and game theory is used to set a reasonable price. In this scheme, each user in the community can perform optimal energy management according to its own situation and protect its own privacy. Li et al.~\cite{li2019design} conduct in-depth research on Singapore's demand-side energy management. They develop a scheme based on smart contract and game theory for price optimization. However, it is worth noting that this scheme takes into account the uncertainties of renewable resources, and uses the receding horizon optimization technique to do demand side management.
Ref. \cite{wu2017m2m} proposes a process optimization scheme based on M2M (Machine-to-Machine) technology. This solution uses smart contracts to publish price records, which are analyzed to get a model of demand response. Similarly, Pop et al.~\cite{pop2018blockchain} adopt smart contracts and incentive mechanism in Ethereum, thereby achieving an accurate demand response mechanism.


Effective demand side management can reduce the negative impact of peak load.
In \cite{wang2019blockchain}, Wang et al.~discuss in detail the possibility of blockchain for demand-side management, and apply smart contracts to achieve load optimization and price optimization. Dang et al.~\cite{dang2019demand} analyze the demand-side management of large industries, and use individual load optimization to achieve optimization for the entire industry.


There exist a number of schemes that focus on the problem of demand response. The work in \cite{erenouglu2020blockchain} explores the feasibility of blockchain to achieve demand response and conducts field experiments. In \cite{Noor2018Energy} and \cite{hajizadeh2020blockchain}, the authors propose demand-side energy management for microgrids.
The model \cite{Noor2018Energy} reduce the peak-to-average ratio based on game theory.
Hajizadeh et al.~\cite{hajizadeh2020blockchain} use a smart contract to provide an automated processing mechanism. The smart contract is negotiated between both parties, so it meets the expectations of consumers and improves the flexibility of the system. 

\subsection{Summary}
We summarize the design goals, targets, and other blockchain-related attributes of existing works discussed in this section, and report the results in Table \ref{management22}.
\begin{table*}[htbp]
	\begin{threeparttable} \footnotesize
		\caption{Blockchain For Intelligent Energy Management}
		\vspace{-0.3cm}
		\label{management22}
		\begin{tabular}{c c c c c c c c c}
			\toprule
			Ref.   & Design goal  & Target   &   Consensus          &   Blockchain      \\
			\specialrule{0em}{1pt}{1pt}
			\midrule
			Wang\textit{ et~al.}\cite{wang2018blockchainassisted} &  Market balance & Crowdsourcing   &  -                 &   Hyperledger   \\
			\specialrule{0em}{1pt}{1pt}
			
			Cutsem\textit{ et~al.}\cite {van2020cooperative} 	&  Demand response &		Smart building			&   PoW 		  	  						&		Ethereum 	\\
			\specialrule{0em}{1pt}{1pt}
			
			Pop\textit{ et~al.} \cite {pop2018blockchain} &		Demand response 	&	Prosumer				&		PoS			 					&		Ethereum	\\
			\specialrule{0em}{1pt}{1pt}
			
			Ioannis\textit{ et~al.}\cite{TsolakisA} &  Demand response 			& Prosumer  &  -			 		 				&		Ethereum 	\\
			\specialrule{0em}{1pt}{1pt}
			
			Edmonds\textit{ et~al.}\cite {edmonds2020blockchain} &		Demand response		&		Smart homes	&		PoW						&		Ethereum	\\
			\specialrule{0em}{1pt}{1pt}
			
			Shao\textit{ et~al.}\cite {shao2019grid}	& Connection freely   	&	Any 				&		iPoW 				  	&       Semi-center 	\\
			\specialrule{0em}{1pt}{1pt}
			
			Tarek\textit{ et~al.}\cite {alskaif2019decentralized} 	&			Flow optimization		&		Prosumer			&		-			 			&		Private	\\
			\specialrule{0em}{1pt}{1pt}
			
			Li\textit{ et~al.}\cite {li2019design} 	&  Flow optimization &	Prosumer				& 	dPoS			  				    &   Bitcoin \\
			\specialrule{0em}{1pt}{1pt}
			
			Dang\textit{ et~al.}\cite {dang2019demand} 	&  Load optimization &	Big industrial				& PoW			  	  			    &  - \\
			\specialrule{0em}{1pt}{1pt}
			
			George\textit{ et~al.}\cite{LazaroiuBlockchain} & 	Distribution & Prosumer &  -					 				&		- 	\\
			\specialrule{0em}{1pt}{1pt}
			
			Rafal\textit{ et~al.}\cite{SkowronskiOn}  &    Distribution  	 &  Any   			&  Hybrid        &     Ethereum  \\
			\specialrule{0em}{1pt}{1pt}
			
			Fan\textit{ et~al.}\cite{FanConsortium}  &     Flexible regulation   & Any   &  dPoS            &     Consortium   \\
			\specialrule{0em}{1pt}{1pt}
			
			Wu\textit{ et~al.}\cite {wu2017m2m} &		Cost optimization	&		Any				&		PoC				 					&		Bitcoin	 \\
			\specialrule{0em}{1pt}{1pt}
			
			Afzal\textit{ et~al.}\cite {afzal2020blockchain} 	&  Cost optimization  &	Smart homes 				& 	 -		  	  			     &   Ethereum \\ 
			\specialrule{0em}{1pt}{1pt}
			
			Yahaya\textit{ et~al.}\cite {yahaya2020blockchain} 	&			Cost optimization 	&		Smart homes		&		PoW									&		Ethereum	\\
			\specialrule{0em}{1pt}{1pt}
			
			\bottomrule
		\end{tabular}
	\end{threeparttable} \normalsize	
\end{table*}
After reviewing the aforementioned blockchain-based solutions for smart grids,
we obtain the following findings:
\begin{itemize}

	\item The decentralized nature of blockchain facilitates efficient energy management in smart grids, including energy distribution and demand side management. Many solutions use blockchain to carry out distributed transformation of the smart grid so that it can adapt to the management of new distributed energy sources. Distributed management can realize the decentralization of management rights and solve the centralized management problems of traditional power grids, which can facilitate the conversion of consumers to prosumers.

	\item Most management solutions that combine blockchain and smart grids can realize effective demand response balance, entire system optimization, distributed storage management of resources, and overall market balance control \cite{patil2021study}.

	\item Some solutions are suitable for various application scenarios and have good universality. There are also solutions that target specific scenarios, such as smart home, prosumer, and outsourcing fields. Blockchain has played an important role in all these schemes.

	\item The multiple combinations of blockchain platforms and consensus algorithms provide a huge number of choices for various application scenarios. At the same time, some solutions even innovatively propose new blockchain platforms and consensus algorithms to perfectly adapt to their needs.

	\item Smart contracts and incentive mechanisms closely integrated with blockchain can effectively realize automated operations of a system, eliminate the intervention of third parties, and attract more users to join. Especially in the management of smart grids, smart contracts can give full play to their greatest advantages, eliminate hidden human hazards, increase system security, and reduce required costs.

    \item However, the biggest problem with most solutions is that they all ignore user privacy protection. They simply mention that their solutions can protect privacy, but do not provide technical details on privacy protection.

\end{itemize}

%% file: 5trading.tex
\section{Blockchain For Energy Trading}\label{trading}

\subsection{Current Issues in Energy Trading}


Trading energy in a competitive energy market can significantly improve energy efficiency,
reduce cost of energy consumers, and increase income of energy providers.
However, because the energy industry is intensive in capital, labor, and resource,
only large-scale energy giants can survive and thus monopolize the energy market.


As the emergence of distributed energy sources such as solar panels, wind turbines,
and electric vehicles, it is important for smart grids to support prosumers, which
may simultaneously be energy consumers and providers. Unfortunately,
the current wholesale energy market only accommodates large-scale electricity plants,
while the prosumers are normally small in capability. Although the retail energy market
may accommodate small-scale prosumers, it is hard to efficiently support so many prosumers.
Hence blockchain is introduced to smart grids
for improving energy efficiency and usage of renewable energy.


In this section, we categorize energy trading from two aspects: peer-to-peer energy trading
and transaction pricing. 
We also provide a summary on existing schemes.

\subsection{Blockchain-Enabled Peer-to-Peer Energy Trading}\label{num3.1}


The first challenge in energy trading comes from the prosumers distributed in a smart grid, because prosumers in traditional power grids do not have sufficient rights, which makes traditional power grids hard to fulfill the current power market requirements \cite{pop2020trading}. On the other hand, as more prosumers use the rooftop solar photovoltaic technology to generate electricity, the surplus electricity of individuals and organizations become more common.
The decentralized nature of blockchain makes it well-suited to solve these problems. Ref. \cite{kumari2020blockchain} develops a blockchain-enabled energy trading platform to realize effective management of renewable resources, meeting the needs of residents, factories and other nodes accordingly.  


The diversity of blockchain platforms also provides space for different design options.
Sabounchi and Wei \cite {dang2019demand} propose a transformation scheme based on Ethereum, which makes each prosumer a participant and manager of the system. At the same time, in order to incentivize participation of prosumers, the authors combine smart contracts to realize fair auction. Plaza et al.~\cite{plaza2018distributed} introduce a blockchain-based solution to serve the energy community sharing solar energy. Like the work in \cite{plaza2018distributed}, Yang \cite{yang2020blockchain} proves the feasibility of implementing distributed trading of renewable energy based on the Ethereum platform.
Moreover, the authors in~\cite{pipattanasomporn2018blockchain} propose to use Hyperledger
for trading surplus electricity and compensating the sellers accordingly. Similarly, Jamil et al. \cite{jamil2021peer} use machine learning and Hyperledger to implement an energy
trading platform that realizes electricity usage forecast and
on-demand distribution. In addition, Kavousi-Fard et al.~\cite{kavousi2021effective} propose a new consensus algorithm named Relaxed Consensus Innovation (RCI), which is suitable for the p2p energy trading market, especially microgrids.


Pee et al.~\cite{pee2019blockchain} make use of blockchain and smart contracts to implement automatic transactions, eliminating trusted third parties and ensuring transparency, immutability and traceability of energy trading.
Similarly, Yu et al.~\cite{yu2018distributed} incorporate smart contracts to realize transparent and independent energy transactions for prosumers. 
Okoye et al.~\cite{okoye2020blockchain} improve the access rules of blockchain and greatly increase the transaction speed. 


In addition to the above solutions for prosumers, there also exist proposals for traditional consumers to achieve supervision and optimization of the transaction system \cite{9121447, liu2020application}. ElectroBlocks \cite{tanwar2020electroblocks} solves the problems of insufficient demand and supply in traditional power grids through cost aware and store aware algorithms. Gao et al.~\cite{GaoGridMonitoring} use a sovereign blockchain to implement the supervision function of a trading system. This solution employs smart contracts to control consumer access, detect malicious nodes, and make transactions transparent. Danzi et al.~\cite {danzi2018blockchain} make use of a layered mechanism based on Ethereum to realize a transparent energy transaction market. They first implement a distributed architecture using blockchain. On this basis, the second blockchain is combined with smart contracts to achieve real-time information acquisition, implement fair pricing, and achieve demand response. The work~\cite{wang2019energy} proposes an optimization model based Hyperledger for operating a crowdsourcing energy system. In order to improve the rationality in energy trading, Wang et al.~present a two stage arithmetic algorithm. Furthermore,  \cite{khorasany2020lightweight}  develops an anonymous location proof algorithm, which employs distance and reputation to determine the most suitable trading partner. Umoren et al.~\cite{umoren2020blockchain} exploits the latest 5G technology and graphical interface to enhance the efficiency and supervision of the trading system. Like \cite{umoren2020blockchain}, ET-DeaL\cite{kumari2020deal} proposed by Aparna Kumari also uses the 5G technology. Additionally, it combines IPFS (InterPlanetary File System) with Ethereum to improve the security and privacy of energy transactions. 

\subsection{Fair Energy Pricing}


The auction platform is of great significance in the pricing of energy transactions. Ref.~\cite{HahnSmart} proposes an auction mechanism based on smart contracts, which achieves reasonable energy pricing and eliminates third parties. More importantly, the mechanism uses Vickrey auctions to further ensure transaction fairness. The scheme presented  in \cite{muzumdar2021trustworthy} also employs the Vickrey auction. It develops an iterative Vickrey–Clarke–Grove method and combines it with the proof-of-stake consensus blockchain to realize an energy trading mechanism with incentive functionality. Wang et al.~\cite{wang2018decentralized, wang2017novel} design a continuous double auction scheme to address the problem of excessive price fluctuations. In order to ensure security of the auction process, the authors use the multi-signature technique and the adaptive attack strategy to ensure the accuracy and stability of the pricing process. In \cite{stubs2020blockchain}, Marius et al. propose a mechanism similar to double auction, which optimizes distribution while effectively reducing the computational load of blockchain nodes. The work in \cite{hassan2021optimizing} analyzes the technical work of developing blockchain energy auctions from a green perspective.


Besides auction pricing, there also exist schemes that determine accurate and effective prices based on energy demand and supply. Ref. \cite{yue2019dynamic} proposes a dynamic pricing scheme for a blockchain-based ecosystem architecture, in which  price is determined by an optimization problem between the microgrid and distributed system operators. This scheme incorporates smart contracts to realize automatic transaction execution. Li et al.~\cite{li2019blockchain} develop a pricing scheme to build a distributed non-cooperative market, which can achieve Nash equilibrium to ultimately maximize the welfare of the entire system. Blockchain and smart contracts are employed in this scheme to establish a transactive energy market. Similar to \cite{li2019blockchain}, the solution in~\cite{thomas2017automation} utilizes smart contracts to realize demand pricing and automated arrangements. However, this scheme provides certain rewards to users who actively adjust the price. Moreover, the solution proposed in~\cite{meng2019decentralized} is based on the information exchanged between prosumers. Prosumers conduct real-time multilateral pricing bids based on their respective power needs and actual conditions. This solution uses smart contracts to reflect the interests of the prosumers and encourage them to participate. In \cite{knirsch2018privacy}, Knirsch et al.~make use of load distribution information of power plants and customers for training and modeling, so as to predict electricity consumption and realize reasonable pricing.

\begin{figure}[htbp]
	\includegraphics[width=0.95\linewidth]{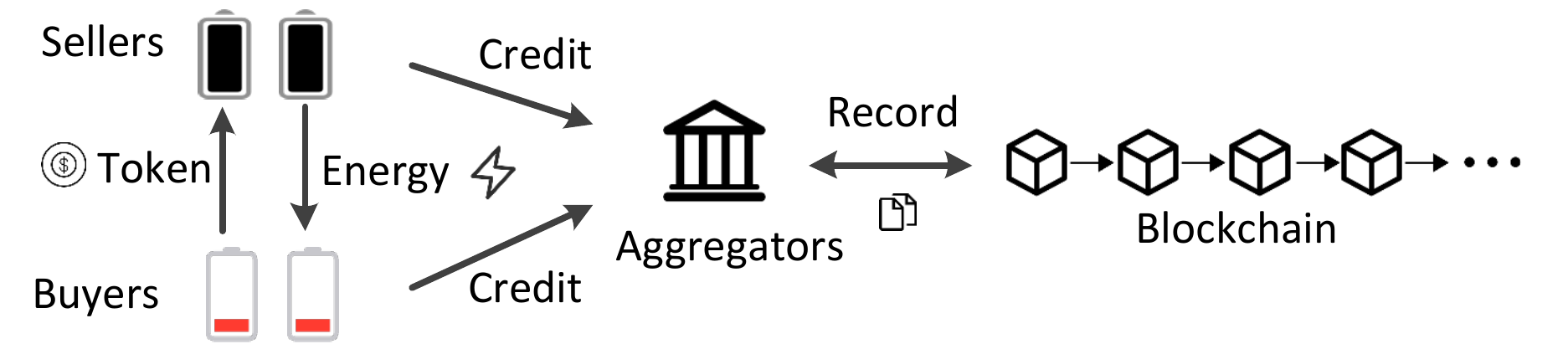}
	\caption{Energy exchange process based on the scheme in \cite{LiConsortium}.}
	\label{fig:trading}
\end{figure}

The combination of blockchain and game theory can achieve fair and effective pricing in the energy trading market. Ref. \cite{LiConsortium} combines consortium blockchain to ensure transaction security, and uses the Stackelberg game to achieve reasonable pricing (shown in Figure. \ref{fig:trading}). Wu et al.~\cite{wu2018game} implement a secure energy trading platform to solve the problem of wind energy heating. The platform is a three-tier structure and has a side-payment mechanism. It is worth noting that the authors employ sequential games to achieve Nash equilibrium, ensuring reasonable and accurate energy pricing. In \cite{guo2020combined, guo2020architecture}, the authors make use of the Stackelberg game to achieve fair pricing between a grid and a communit, thyereby obtaining the maximum benefit. In these works, the blockchain platforms mainly serve to guarantee security and privacy. Smart contracts can effectively improve transaction efficiency. Guo et al.~\cite{guo2020architecture} design a new Byzantine consensus to improve the success rate. Doan et.al~\cite{doan2021peer} exploit a double auction-based Stackelberg game to encourage individual users to participate in energy trading and achieve maximum social welfare.


In addition to the above-mentioned methods, pricing in the energy trading process is also applied to certain special scenarios. Chen et al.~\cite{chen2019framework} propose a pricing algorithm for end devices in edge computing, and blockchain is used to realize distributed transactions. In \cite{wang2020research}, Wang et al.~apply smart contracts to implement a bidding system for V2G. This system employs a particle swarm to analyze and model the smart grid, so as to realize rational bidding. The authors also develop genetic algorithms to further optimize the bidding results and maximize the benefits of the entire system.

\subsection{Summary}

We summarize the key technologies and blockchain-related attributes of the works mentioned above in this subsection. The details are shown in Table \ref{trading22}.
\begin{table*}[htb]
	\begin{threeparttable} \footnotesize
		\caption{Blockchain For Energy Trading}
		\vspace{-0.3cm}
		\label{trading22}
		\begin{tabular}{c c c c c c c c}
			\toprule
			Ref.   & Target   &   Consensus    &    Demand res.    
			&    Contract     &   Blockchain       \\
			\specialrule{0em}{1pt}{1pt}
			\midrule
			Gao\textit{ et~al.}\cite {GaoGridMonitoring} &  Any   &  - &   No  
			&      Used     &     Sovereign    \\
			\specialrule{0em}{1pt}{1pt}
			
			Okoye\textit{ et~al.}\cite{okoye2020blockchain} &  Microgrid   &  PoS &   No 
			&      Used     &     Consortium      \\
			\specialrule{0em}{1pt}{1pt}
			
			Yu\textit{ et~al.}\cite{yu2018distributed} &  Any  &  -			&		Yes	
			&	 Used 			&		Ethereum 	  		\\	
			\specialrule{0em}{1pt}{1pt}
			
			Manisa\textit{ et~al.}\cite {pipattanasomporn2018blockchain}  &  Any  &  PoC				&	No	
			&	 Used 			&	Hyperledger 		\\
			\specialrule{0em}{1pt}{1pt}
			
			Wang\textit{ et~al.}\cite {wang2019energy}			&	Any 				&		RBFT 		&		Yes	
			&   	Used  &   Hyperledger			 \\
			\specialrule{0em}{1pt}{1pt}
			
			Umoren\textit{ et~al.}\cite{umoren2020blockchain}			&		V2G			&		-	&		Yes		
			&			Unused		&		Private			\\
			\specialrule{0em}{1pt}{1pt}
			
			Khorasany\textit{ et~al.}\cite{khorasany2020lightweight}			&		Any			&	A-PoL	&		Yes	
			&			Used		&		Lightweight		\\
			\specialrule{0em}{1pt}{1pt}
			
			Meng\textit{ et~al.}\cite{meng2019decentralized}			&		Prosumer			&		-	&		Yes		
			&			Used		&		Ethereum			\\
			\specialrule{0em}{1pt}{1pt}
			
			Wang\textit{ et~al.}\cite{wang2018decentralized}			&		Microgrid			&		-	&		No	
			&			Used		&		Bitcoin		\\
			\specialrule{0em}{1pt}{1pt}
			
			Hahn\textit{ et~al.}\cite {HahnSmart} 	&		Any				&		PoC			&		Yes	
			&   Used   &   Ethereum			  \\
			\specialrule{0em}{1pt}{1pt}
			
			Wu\textit{ et~al.}\cite {wu2018game} 	&	North China				&			-		&		No	
			&   Used   &   Ethereum			   \\
			\specialrule{0em}{1pt}{1pt}
			
			Danzi\textit{ et~al.}\cite {danzi2018blockchain}	&	Any 				&		PoW			&		Yes		
			&			Used			&		Ethereum						\\
     		\specialrule{0em}{1pt}{1pt}
			
			Li \textit{ et~al.}\cite {LiConsortium} &		IIoT			&		PoW				&	No		
			&			Unused		&		Consortium		\\ 
			\specialrule{0em}{1pt}{1pt}
			
			Yue\textit{ et~al.}\cite {yue2019dynamic}  		&	Prosumer 				&		- 		&		Yes 
			&   	Used  &   -				 \\
			\specialrule{0em}{1pt}{1pt}
			
			Li\textit{ et~al.}\cite {li2019blockchain}	&	Any					&		-				&		No		
			&			Used		&	 Ethereum 			\\
			\specialrule{0em}{1pt}{1pt}
			
			Guo\textit{ et~al.}\cite {guo2020combined}  &	Any			&		PoW				&	No		
			&			Unused		&		Bitcoin		\\
			\specialrule{0em}{1pt}{1pt}
			
			Wang\textit{ et~al.}\cite {wang2020research}	&	Prosumer & - &  No   
			& 		Used 		&		Ethereum   \\
			\specialrule{0em}{1pt}{1pt}
			
			Fabian\textit{ et~al.}\cite {knirsch2018privacy}	&	Any 				& 	PoW 		
			&			No 				&   Used      &   -   \\
			
			\bottomrule
		\end{tabular}
		
	\end{threeparttable} \normalsize
\end{table*}
Having reviewed the aforementioned blockchain-based solutions for smart grids,
we obtain the following findings:
\begin{itemize}
	\item Most of the schemes that integrate blockchain with energy trading can be applied to any transaction scenario. A few schemes are specifically developed for V2G, microgrids, prosumers, and industrial IoTs.
	
	\item Blockchain as a carrier of the cryptocurrency or token has the desirable financial property. This makes blockchain well-suited for payment processes. More importantly, the decentralized nature of blockchain makes it both fault tolerant and highly secure. These features make it an ideal choice as a platform for energy trading in smart grids \cite{lotfi2020transition}. 
	
	\item In the application scenario of energy trading, the incentive mechanism of blockchain is particularly important. It has a positive effect on promoting energy trading and reasonable pricing. Due to the high requirements for reasonable pricing of energy transactions, current schemes also combine game theory, dynamic pricing, and auction technologies on the basis of incentive mechanism design.
	
	
	\item The consensus mechanism can further ensure security and efficiency of the transaction process. Smart contracts can be employed to realize complex payment processing logics, such that payment can be made automatically and securely according to predefined rules.
	
	\item In terms of security, we find that some schemes cannot avoid the security and privacy risks that could occur during the energy trading process. Additionally, due
to the delay of blockchain itself, energy demand requests may not be responded in real time, which significantly affects the effective distribution of energy.

\end{itemize}

In addition, security and privacy issues are very important in energy trading, which are to be discussed in the next section, while energy trading in microgrids is covered in Sec.~\ref{Microgrid}.

%% file: 6security-and-privacy.tex
\section{Security And Privacy Protection}\label{security}

\subsection{Current Issues in Smart Grid Security and Privacy}


A smart grid is an important infrastructure, so its security and privacy protection are the prerequisites for high-confidence system operations.
In recent years, power grids face serious security risks, threats and various attacks, including but not limited to:smart meter attacks, time synchronization attacks, physical network attacks, denial of service (DOS) attacks, false data injection attacks, replay attacks, double spend attacks, network attacks and GPS spoofing attacks.		


Traditional schemes combine Kalman filter, stochastic game, sparse optimization and other technologies to prevent and detect attacks. However, most of these schemes can only be targeted at specific situations thus cannot resist all attacks. Furthermore, most of the traditional schemes are based on the centralized power grids, which can not adapt to the new distributed ones.
Blockchain, as a transparent distributed ledger, can adapt to the new smart grids. In addition, the mining mechanism, consensus mechanism, and the avoid of third parties of blockchain make it an effective way to resist most of the attacks and provide a safer environment.


In this section, we summarize the related works from the perspectives  of security and privacy enabled by blockchain. We analyze each category comprehensively and present our insights obtained from this study in the end. 

\subsection{Blockchain for Smart Grid Security}\label{num1}


In the process of transition to a distributed structure, the energy market is prone to potential security risks and crisis of confidence.
In \cite {hua2019blockchain}, Hua and Sun propose a transaction program for energy and carbon markets. In order to further improve security and robustness of the system, the authors adopt implement the so-called ``payment-to-public-key-hash'' with multi-signatures. Abdella et al.~\cite{abdella2019architecture} present a three-layer model to address the distributed energy management problem. The three-layer structure works in conjunction with each other to effectively improve security and efficiency. Ref. \cite{tanaka2017blockchain} highlights the important role of digital grid routers in smart grids, and proposes to integrate blockchain with digital grid routers to build a safer decentralized energy exchange platform.
The work in~\cite {MannaroCrypto} describes the method to change Sardinia's power grid in response to security risks caused by distributed transformation. This project implements a trading system in which blockchain acts as a middleman for the transaction. When a producer produces energy units, the smart meter sends the message to blockchain. Smart contracts automatically realize the interaction between information and tokens. There is also an agent in the system to help the sellers formulate strategies to maximize the seller's interests and facilitate rational use of resources.



Traditional power grids often rely on reliable third parties, which generally brings unknown security risks to the system. EnergyChain \cite{AggarwalEnergyChain} provides a solution specifically for smart homes without third parties.


External attacks are one of the common problems that cause security risks.
In \cite {ferrag2019deepcoin}, Ferrag and Maglaras propose an energy framework called DeepCoin based on blockchain and deep learning. The authors adopt a practical Byzantine fault-tolerant algorithm to improve efficiency. In order to counter against the cyber-attacks, different solutions have been proposed in \cite{mylrea2017blockchain, LiangDistributed}. The program in \cite{mylrea2017blockchain} simulates a power grid system with limited resources. Liang et al.~\cite{LiangDistributed} conduct experiments based on IEEE-118 and smart contracts to prove the effectiveness of the schemes. In order to enhance privacy protection and power
security, Kaur et al. \cite{kaur2021blockchain} propose a framework based on
software-defined networking and blockchain. The framework
combines a secure and efficient mutual authentication protocol
based on the elliptic curve cryptosystem and a smart contract
for demand response. For fake data injection attacks, Samy
et al. \cite{samy2021towards} develop an abstract model based on blockchain.

\begin{figure}[htbp]
	\includegraphics[width=.95\linewidth]{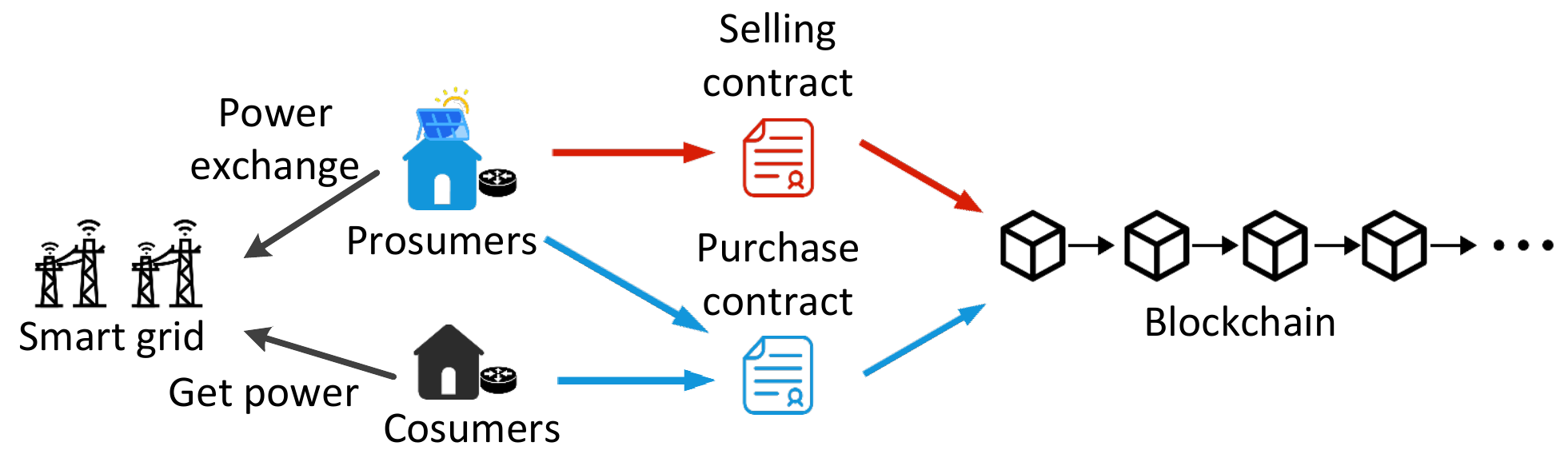}
	\caption{The workflow of the scheme in \cite{shuaib2018using}. }
	\label{fig:security}
	
\end{figure}

On the other hand, 
many solutions have been optimized for secure payment and access for the energy market. 
Researchers make effort to design secure payment platforms for energy exchange~\cite{luo2018distributed, shuaib2018using}.
The work in \cite{luo2018distributed} proposes a multi-layer grid architecture for consumers. The authors implement a trusted settlement mechanism at the second layer. Shuaib at al.~\cite{shuaib2018using} use smart contracts to verify the possibility of blockchain as a secure payment for distributed energy (shown in Figure. \ref{fig:security}).
Except the payment mechanism, Garg et al.~\cite{GargAn} develop a secure access mechanism for V2G by using elliptic curve cryptography. This scheme can ensure strong anonymity and low cost. Ouyang et al.~\cite{ouyang2017preliminary} focus on direct transactions for large scale power consumers. The framework adopts smart contracts to implement access and settlement mechanisms to ensure the security of transactions. 
Bera et al. \cite{bera2021designing} propose
a new access control protocol named DBACP-IoTSG, which
is implemented in a private chain and has the characteristics
of low cost and high security.
Alcaraz et al.~\cite{alcaraz2020blockchain} consider the security problems caused by the high coupling degree of the entities, and propose a three-tier architecture to reduce the coupling degree. Tang et al. \cite{tang2021multiauthority} propose an
anonymous authentication scheme based on ring signatures,
which can realize tracking without exposing user information.

In \cite{jiang2019blockchain} and \cite{huang2019development}, two solutions are mainly proposed for smart grids as an IoT system. While ensuring security, Jiang et al.~\cite{jiang2019blockchain} use smart contracts and DPoS consensus to improve efficiency. Huang at al.~\cite{huang2019development} employ Sigfox and Ethereum to confirm the feasibility of the solution. The work in \cite{sikeridis2020blockchain} mainly describes the security risks in the energy transmission system and the corresponding solutions making use of the  decentralization, transparency and immutability properties of blockchain. Moreover, Ref. \cite{inayat2018load} discusses possible security issues in the AMI (Advanced Metering Infrastructure) and proposes a framework that can achieve load balancing. This framework integrates the blockchain incentive mechanism and the encryption/signature algorithm to ensure the security of the energy market and improve the efficiency of demand response. It is worth mentioning that BBARS \cite{wang2019bbars} employs a variety of public key encryption technologies to achieve the security of V2G's anonymous rewards.


Unlike the previous schemes, He et al.~\cite{he2020joint} discuss the methods to combine the advantages of the photovoltaic power and carbon market.
This scheme provides a dual-blockchain structure, in which the two chains cooperate and supervise each other to achieve data sharing. The authors also prove the usability on the Ethereum platform.

\subsection{Blockchain for Power Grid Privacy}\label{num1.2}


One of the reasons for privacy leaks lies in outside attacks or the introduction of dishonest third parties.
It is pointed out that many solutions use blockchain to transform a smart grid, but the public ledger may lead to user privacy leakage due to malicious attacks \cite{GaiPrivacy}.
This work adopts a consortium blockchain, where each energy transaction intermediary acts as a miner node, and the transaction intermediary is composed of energy reserve equipment and token banks.
It also uses smart contracts to design a black box and create a virtual account for each user.
Chen et al.~\cite{chen2020blockchainPrivacy-Preserving} employ group signatures based blockchain to build a platform, which can specifically solve the privacy leakage problems of WSNs (Wireless Sensor Networks).
In \cite {li2020blockchain}, Li et al.~propose FeneChain that makes use of anonymous authentication to strengthen privacy protection. The authors design a mechanism based on time commitment to ensure fairness and accuracy of energy transactions. Lu et al.~\cite {lu2019secure} incorporates a reputation-based fairness proof mechanism, reducing the user's computational burden and lowering the system's entry barrier. The scheme divides the electricity market transaction model into two levels that can cooperate with each other and provide a secure environment.


Ref. \cite{gai2019permissioned} points out that it is easy to launch attacks at the edge layer. Therefore, this work intends to use edge computing to enhance the security of edge nodes. At the same time, encrypted channels are introduced to ensure secure communications between edge devices. Blockchain can achieve decentralization to ensure the authenticity of the data, and users use pseudonyms to participate in transactions and preserve their privacy. Smart contracts are adopted to achieve the most efficient energy distribution strategy. In response to the privacy issues of edge devices, Guan et.al~\cite{guan2021blockchain} propose the BPM4SG scheme, which combines techniques such as multi-party secure computation, smart contracts, and ring signatures. The experimental results prove that this scheme has good performance advantages.

\begin{figure}[htbp]
	\includegraphics[width=0.95\linewidth]{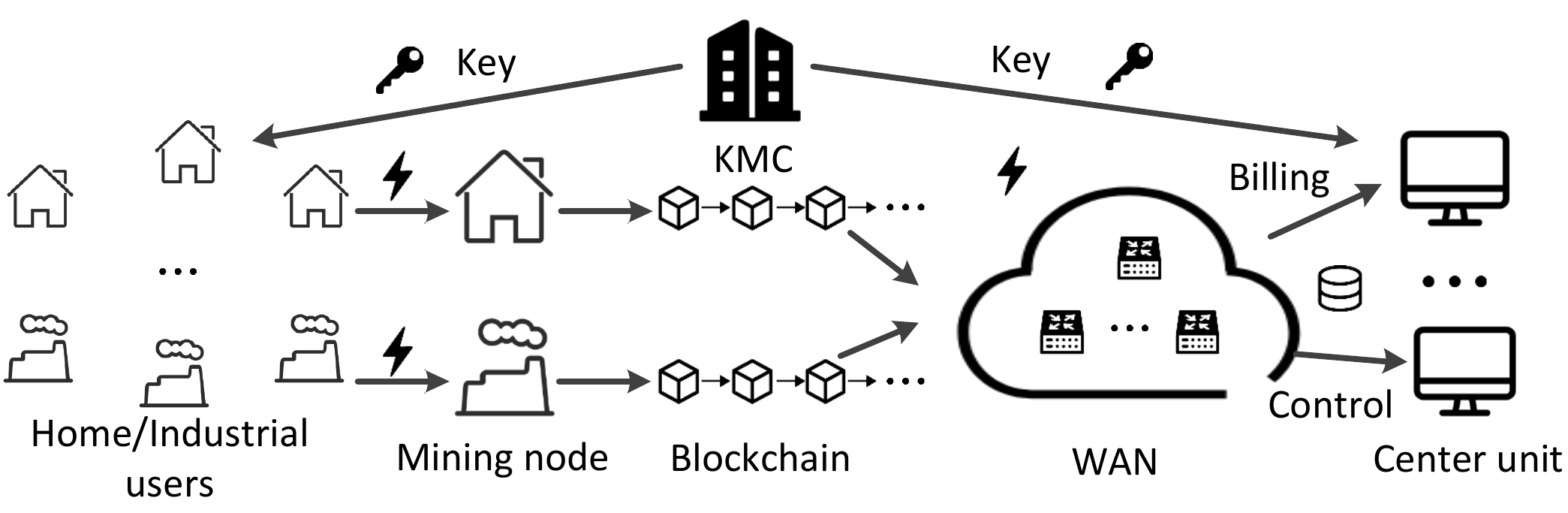}
	\caption{ The description of the key steps of the scheme in \cite{guan2018privacy}.}
	\label{fig:guan2018privacy}
\end{figure}

Sestrem et al.~\cite{sestrem2020cost} mainly tackle the potential privacy risks in the optimization process with blockchain. They protect the privacy of the entire system by establishing sidechains. Samuel et al.~\cite{OmajiA} propose an admission control mechanism based on Ethereum to solve the dilemma of service provision and privacy disclosure. The authors also combine Pagerank to encourage consumers to join. The work in \cite{guan2018privacy} applies the method of user group management to address the privacy leakage problem for smart meters. It establishes private blockchains to record the data of each group, and uses multiple pseudonyms to further protect the privacy of users (shown in Figure. \ref{fig:guan2018privacy}). Li et al. \cite{li2021lightweight} develop a privacy protection scheme in the process of price optimization,
which is based on the secure signature authentication mechanism
and identity-based proxy re-encryption strategies. In response to the privacy challenge of data aggregation, Singh et.al~\cite{singh2021blockchain} present a scheme called BHDA, which combines deep learning and homomorphic encryption to improve the performance of data aggregation and privacy protection with minimal computational overhead. Lu et.al~\cite{lu2021edge} propose a solution named EBDA to address the privacy issues caused by data aggregation , which integrates not only blockchain, but also edge computing technologies.

In the process of decentralization of smart grids, privacy issues are also extremely easy to arise.
In \cite{deng2019real} and \cite{son2020privacy}, the authors aim to achieve privacy protection for prosumers. Deng et al.~\cite{deng2019real} use a central operator to summarize the data, preventing prosumers from obtaining data not belonging to them. The proposed scheme also presents a new primal-dual gradient method for rational energy scheduling. Son et al.~\cite{son2020privacy} make use of functional encryption and smart contracts to improve system security and protect privacy. Lu et al.~\cite{lu2019blockchain} propose a solution based on SDN (Software Defined Network) technology. Experiments prove that this solution has good accuracy. Keshk et al.~\cite{keshk2019privacy} develop a two-level model. The privacy module is mainly an enhanced blockchain based on proof-of-work. It is used to ensure data integrity and mitigate data poisoning attacks. The anomaly detection module employs deep learning to integrate data. The two modules cooperate with each other, which greatly increases the security and efficiency.  Zhang et al.~\cite{zhang2020privacy} employ blockchain and ring signature to enhance privacy protection. The innovation of this work lies in its adoption of the point multiplication of elliptic curve, which greatly reduces the system complexity. The work in \cite{liu2021privacy}
demonstrates a privacy-preserving network protocol based
on multi-level encryption. Guan et al. \cite{guan2021achieving} provide a general
model based on ciphertext-policy attribute-based encryption and
credibility-based equity proof consensus, which can achieve
fine-grained access control through transaction arbitration in
the form of ciphertext.

\subsection{Summary}
We summarize the design goals, key technologies, and other blockchain-related attributes of the existing solutions mentioned above. The results are reported in Table \ref{security22}.

\begin{table*}[htb]
	\begin{threeparttable} \footnotesize
		\caption{Blockchain For Security And Privacy Protection}
		\vspace{-0.3cm}
		\label{security22}
		\begin{tabular}{c c c c c c c c c}
			\toprule
			Ref.  & Design goal    &   Consensus          &    Contract     &   Blockchain   \\
			\specialrule{0em}{1pt}{1pt}
			\midrule
			
			Luo\textit{ et~al.}\cite{luo2018distributed}	  &   Secure payment	  			&  -
			&      Used     &     Double-chain      \\
			\specialrule{0em}{1pt}{1pt}
			
			Garg\textit{ et~al.}\cite{GargAn}			&  Security access				&		-
			&			Used		&		-				\\
			\specialrule{0em}{1pt}{1pt}
			
			Huang\textit{ et~al.}\cite{huang2019development}		&		Wireless security							&		-			
			&			Unused			&		Ethereum				\\
			\specialrule{0em}{1pt}{1pt}
			
			He\textit{ et~al.}\cite{he2020joint}			&    Market integration					&		Hybrid 		
			&   	Unused  &   Double-chain			  \\
			\specialrule{0em}{1pt}{1pt}
			
			EnergyChain~\cite{AggarwalEnergyChain}	&  Third party   & PoW
			& 		Unused 		&		EnergyChain     \\
			\specialrule{0em}{1pt}{1pt}
			
			Gai\textit{ et~al.}\cite{GaiPrivacy}	&  Third party					& 	PBFT
			&   Unused      &   Consortium     \\
			\specialrule{0em}{1pt}{1pt}
			
			Gai\textit{ et~al.}\cite {gai2019permissioned}   &  Service optimization  				& 	 PoW			
			&   Used      &   Permissioned   \\
			\specialrule{0em}{1pt}{1pt}
			
			Guan\textit{ et~al.}\cite {guan2018privacy}  &  Service optimization 			& 	 PoW			
			&   Unused     &   Private  
			\\
			\specialrule{0em}{1pt}{1pt}
			
			Lu\textit{ et~al.}\cite{lu2019blockchain}	&   Distributed design  					&		-		
			&			Unused		&		-			\\
			\specialrule{0em}{1pt}{1pt}
			
			Hua and Sun~\cite{hua2019blockchain} &   Distributed design      &  -
			&      Used     &     -      \\
			\specialrule{0em}{1pt}{1pt}
			
			Juhar\textit{ et~al.}\cite {abdella2019architecture}	&    Distributed design 			&		PoS
			&   Used   &   Hybrid			    \\
			\specialrule{0em}{1pt}{1pt}
			
			Deng\textit{ et~al.}\cite{deng2019real}		&   Distributed design				&   -
			&	 Unused 			&   -  \\
			\specialrule{0em}{1pt}{1pt}
			
			Keshk\textit{ et~al.}\cite {keshk2019privacy} 	&   Distributed design 			& 	 ePoW			
			&   Unused     &   -	  \\
			\specialrule{0em}{1pt}{1pt}
			
			Son et al.~\cite{son2020privacy} &   Distributed design 	&   -  
			&  Unused   		&		-     \\
			\specialrule{0em}{1pt}{1pt}
			
			DeepCoin~\cite{ferrag2019deepcoin} 	&	Anti-attack				&		PBFT				
			&			Unused		&	 - 							\\
			\specialrule{0em}{1pt}{1pt}
			
			Li\textit{ et~al.}\cite{li2020blockchain}	 	& Anti-attack		&		PBFT			
			&		Used		&		Consortium			\\
			
			\bottomrule
		\end{tabular}	
	\end{threeparttable} \normalsize
\end{table*}

Having reviewed the aforementioned blockchain-based solutions for smart grid security and privacy,
we obtain the following findings:
\begin{itemize}
\item The research works discussed in this section have greatly improved security and privacy of smart grids. The transparency, immutability, and anonymity of blockchain properties can help to effectively protect the supervision, monitoring and management of smart grids. It is worth mentioning that security, as a basic attribute, is reflected in all fields such as wireless networks, smart homes, smart communities, V2G, and IIoT.

\item Blockchain-based solutions can effectively address various security problems caused by distributed transformation, third parties, malicious attacks, and other reasons.

\item Different encryption technologies such as asymmetric encryption, attribute-based encryption, and functional encryption, are employed to enhance security and privacy for smart grids and their users. Blockchain provides the basis for the implementation of these encryption algorithms.

\item It can be seen that the multi-level and multi-chain information transfer model can be realized on the basis of blockchain, and the protection of information privacy can be enhanced accordingly. We believe that it is indeed feasible to design solutions based on blockchain to protect user privacy.

\end{itemize}

%% file: 7microgrid.tex
\section{Blockchain for Microgrid Management}\label{Microgrid}

\subsection{Current Issues in Microgrid}


The concept of microgrid was first proposed by Lasseter. Although there is no formal standards for a microgrid system, this concept has been accepted by the communities and considerable research works have been dedicated to microgrids. We define a microgrid to be a small- to medium-sized and low-voltage power grid that combines various distributed power generations to provide electrical energy for local loads. A microgrid can improve the reliability of power supply on the demand side. 
From the scale perspective, it is a small-scale power grid structure. But from the functionality perspective, a microgrid is an autonomous system that can realize self control, protection, and management, similar to a large-scale power grid. 
However, a microgrid suffers from
high management cost and low revenue, so it can neither meet the requirements of investors nor guarantee trust for consumers.


As distributed and storage systems, various types of blockchains have been integrated with large-scale power grids. It is also possible to apply blockchain to microgrids. Recent years, many proposals based on the combination of microgrid and blockchain have been proposed, tackling issues such as security, power loss issues, automation and energy distribution in microgrids.

\subsection{Security Design of Microgrids}\label{num4.1}


A microgrid must first ensure that it can resist malicious attacks and threats from third-party organizations.
The solution in \cite{di2018energy} is based on Tendermint, which eliminates the dependence on third-party organizations. Mbarek et al.~\cite{mbarek2020enhanced} employ removable software to detect FDI (False Data Injection) attacks.
Different from the above schemes, \cite{BanksBlockchain} intends to solve the potential MIM (man-in-the-middle) attacks and DS (Data Spoofing) attacks in microgrids.
Based on Hyperledger, this work develops a blockchain-based system for sharing status and control information between microgrids. It can maintain dynamic stability for electricity generation, transmission, and distribution. Blockchain provides consensus nodes in the system to  achieve trustworthy data communications between microgrids.

To enhance security and prevent cyber attacks for microgrids, Wang et al.~\cite{wang2019cybersecurity} use DAG (Directed Acyclic Graph) to improve traditional microgrids by combining the data recovery technology and the UT (Unscented Transform) technique for better efficiency and accuracy.
Like \cite{wang2019cybersecurity}, \cite{dabbaghjamanesh2019networked} also uses the UT technique to model uncertainties in demands and electricity supplies. The blockchain enabled IoT scheme can reduce operation cost and enhance security.

In order to further improve privacy protection of a microgrid, Zhang et al.~\cite {zhang2019privacy} make use of a consortium blockchain and the CDA (Continuous Double Auction) mechanism to reduce cost and improve transaction efficiency. They combine FBS (Fair Blind Signature) and the SSS (Secret Sharing Scheme) based $(t, n)$-threshold scheme to improve security and traceability of the system.
For the purpose of detecting faults and preventing damages due to fault damage propagations in a microgrid, Bayati et al.~\cite{bayati2020blockchain} propose a protection technique based on  blockchain. This scheme employs encryption to protect differential relays from cyber attacks.

The security protection of a microgrid is also reflected in the field of finance. In \cite {jeon2019study}, Jeon et al.~use a hybrid blockchain to solve the hidden danger caused by the problem of double spending. And the work in \cite{warutai2019optimal} takes advantage of transparency and immutability of blockchain to implement an accounting model suitable for renewable energy. Experimental results show that it can promote the development of distributed energy resources in the future.


In \cite{AlamTowards}, security problems caused by one-way communications are addressed, where consumers can only accept passively and cannot respond their demands. Alam et al.~propose a double-chain structure to optimize the microgrid along with a negotiation algorithm. The algorithm generates smart contracts and accomplishes negotiations, thereby promoting the optimal transactions of credit.

\subsection{Power Losses of Microgrids}\label{num4.2}


One problem of the traditional power grid is power losses. Because a power station is usually far away from its users, it leads to a large amount of electricity loss during the transmission process. In a microgrid, power loss problems can be effectively alleviated because the microgrid is an energy ecosystem within a small area with a short transmission distance. However, the increasing number of microgrids brings other power loss problems \cite {sanseverino2017blockchain, hernandez2005fuel}. Power losses may also be resulted from issues such as old machines, coal consumption for power supply, and the superposition of energy transactions \cite{di2018technical}. We investigate the research works based on blockchain and find that it is indeed feasible to use blockchain-based solutions for solving the power loss problem in a  microgrid.


With the deep promotion of distributed microgrids and the massive use of renewable energy sources, the power loss problem of renewable energy sources have become more serious. 
Based on blockchain and smart contracts, Danzi et al.~\cite{danzi2017distributed} tackle the power loss problem caused by the high penetration rate of distributed energy resources. For renewable energy, the work in \cite {khalid2020blockchain} explores the challenges facing Pakistan's power imbalance, and proposes an anti-power solution based on Ethereum. Shishkov et al.~\cite{shishkov2019microgrid} employ blockchain to address the power system operation problem of Russian Federation. 


Di Silvestre et al.~\cite {di2019ancillary} propose a comprehensive framework of microgrids. This work considers the issue of providing auxiliary services. In the proposed scheme, each photovoltaic generator  plays a role in voltage regulation, and blockchain can increase the security of the system. Experimental results show that the scheme can effectively improve the voltage regulation, thereby enhancing the robustness of the system.

\subsection{Microgrid Automation}\label{num4.3}

\begin{figure}[htbp]
	\includegraphics[width=.95\linewidth]{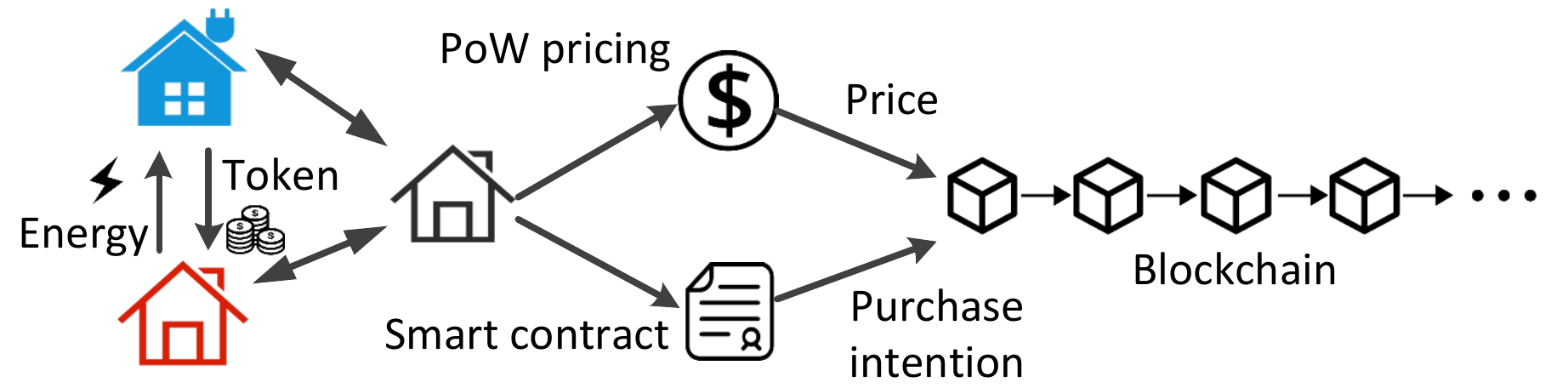}
	\caption{ Microgrid energy interaction based on the scheme in \cite {KangA}. }
	\label{fig:mic}
\end{figure}

Kang et al.~\cite {KangA} use PoW and smart contracts to implement a smart home system based Ethereum. It aims to automate the processing of renewable energy and eliminates the dependence of third parties (shown in Figure. \ref{fig:mic}).
Similarly, Afzal et al.~\cite{afzal2019blockchain} propose a microgrid solution for smart communities, where smart contracts can realize automation for the system.

In \cite{xue2017blockchain}, the authors mainly address the complexity of the traditional power grid nanagement process and the high labor cost. The scheme proposed in Kounelis et al.~\cite{kounelis2017fostering} is committed to solving the problems of high cost of microgrid systems and low user profits. Both solutions \cite{xue2017blockchain, kounelis2017fostering} use smart contracts to achieve automated operations, while greatly reducing operating costs. Furthermore, Gazafroudi et al.~\cite{gazafroudi2020islanded} propose an automatic payment scheme that avoid labor and third parties, thus greatly reducing cost. However, it is not suitable for large-scale application scenarios.

\subsection{Power Distribution of Microgrid}\label{num4.4}



\begin{figure}[htbp]
	\includegraphics[width=.95\linewidth]{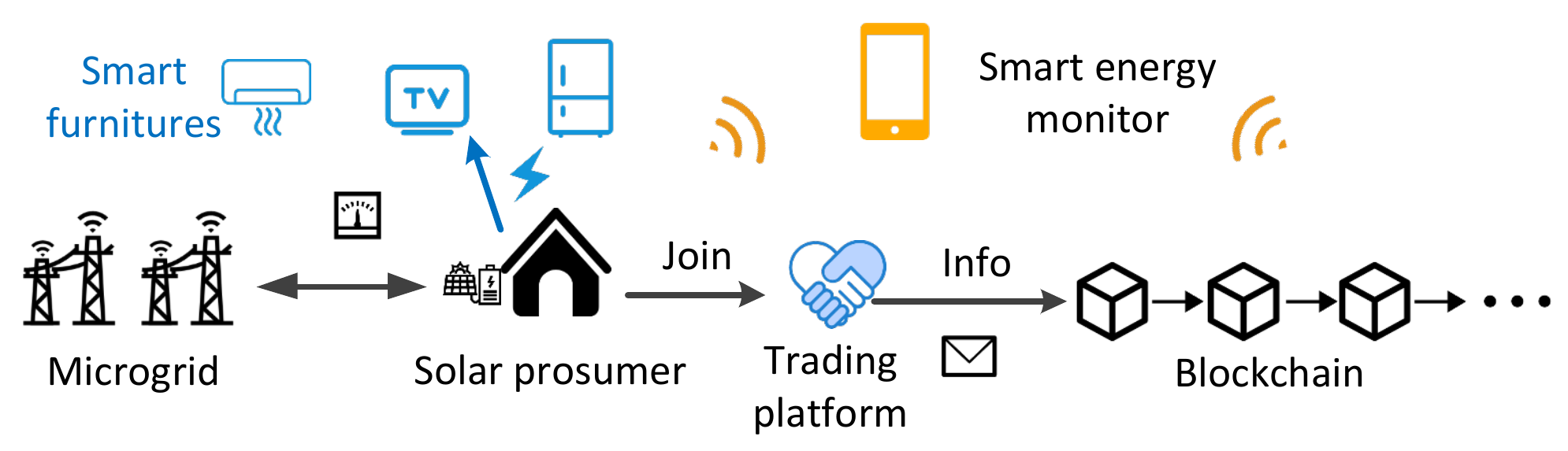}
	\caption{Illustration of the scheme in \cite{AlessandraSmarter} in the field of microgrid.}
	\label{fig:mic-dis}
\end{figure}

Similar to the traditional large-scale power grids, the energy distribution of a microgrid is optimized to meet demand response, improve service quality, and optimize price. 
Alessandra et al.~\cite{AlessandraSmarter} intend to optimize the microgrid process to provide better services (shown in Figure. \ref{fig:mic-dis}).

Like \cite{AlessandraSmarter}, \cite{li2019decentralized} meets the needs of users, and proposes a demand response mechanism for IoT (Internet of Things). This work employs game theory to achieve a more reasonable energy distribution and price.
Li and Nair \cite {li2018blockchain} propose a blockchain based system to generate and trade power, which can optimize price and cost in microgrids. It employs LSB (Linear Supply Bidding) to prove the equilibrium of the electricity market competition. The authors also introduce a DSB (Distributed Supply Bidding) algorithm to achieve reasonable pricing.
Dimobi et al.~\cite {dimobi2020transactive} analyze several  energy market coordination schemes and propose an energy exchange scheme based on Hyperledger Fabric, which can increase the enthusiasm of users for energy investment and optimize cost.


How to achieve optimal power flow in microgrids has also attracted interests of researchers. The work in~\cite{munsing2017blockchains} uses the ADMM algorithm to ensure operational constraints and fair payment without relying on third-party agencies. Blockchain and smart contracts are employed to verify the effectiveness of the solution. With the smart contract technology, a power grid can run automatically and realize automatic payment of transactions.
Van Leeuwen et al.~\cite {van2020integrated} also incorporate the ADMM algorithm to optimize the system's power consumption. The solution utilizes smart contracts and eliminates third parties. In this scheme, the smart contract helps to distribute data to all participants and execute the ADMM algorithm.

\subsection{ Summary }

We summarize the  key technologies and blockchain-related attributes of the existing works in this subsection. The specific results are presented in Table \ref{micogrid22}.
\begin{table*}[htb]
	\begin{threeparttable} \footnotesize
		\caption{Blockchain For Micogrid Management}
		\vspace{-0.3cm}
		\label{micogrid22}
		\begin{tabular}{c c c c c c}
			\toprule
			Ref.  
			&   Consensus  &    Incentive 
			&   Blockchain     &   Key technique    \\
			\specialrule{0em}{1pt}{1pt}
			\midrule
			
			Alam\textit{ et~al.}\cite{AlamTowards} 
			&  - & No
			&     Double-chain    &    -     \\
			\specialrule{0em}{1pt}{1pt}
			
			Jeon\textit{ et~al.}\cite {jeon2019study}
			&   -		&	No
			&		Hybrid 		&	Plasma trading 		 \\
			\specialrule{0em}{1pt}{1pt}
			
			Wang\textit{ et~al.}\cite {wang2019cybersecurity}
			&  -		&	Yes
			&		Private 		&	Unscented trans., DAG			\\
			\specialrule{0em}{1pt}{1pt}
			
			Zhang \textit{ et~al.}\cite{zhang2019privacy}	 	
			&  -  & No
			&     Consortium    &    Blind signature   \\
			\specialrule{0em}{1pt}{1pt}
			
			Danzi\textit{ et~al.}\cite {danzi2017distributed}
			&	PoW	 &  Yes
			&		Ethereum		&  Fairness Control		 		\\
			\specialrule{0em}{1pt}{1pt}
			
			Silvestre\textit{ et~al.}\cite {di2019ancillary}
			&   -	&	Yes
			&		Tendermint 		& Glow-worm swarm 		 \\
			\specialrule{0em}{1pt}{1pt}
			
			Khalid\textit{ et~al.}\cite {khalid2020blockchain} 
			&		PoA		&	
			&		Ethereum		&  Asymmetric crypto.				\\
			\specialrule{0em}{1pt}{1pt}
			
			Kang\textit{ et~al.}\cite{KangA}
			&		PoW	 &  No
			&		Ethereum		&  -					\\
			\specialrule{0em}{1pt}{1pt}
			
			Kounelis\textit{ et~al.}\cite {kounelis2017fostering} 	
			&  PoW  & Yes
			&     Ethereum    &     Helios Coin  \\
			\specialrule{0em}{1pt}{1pt}
			
			Xue\textit{ et~al.}\cite {xue2017blockchain}  
			&  PoW &  No
			&     Ethereum    &    -  &   - \\
			\specialrule{0em}{1pt}{1pt}
			
			Afzal\textit{ et~al.}\cite {afzal2019blockchain}
			&	-		&  Yes
			&   Ethereum				&    Branch and Bound    \\
			\specialrule{0em}{1pt}{1pt}
			
			Alessandra\textit{ et~al.}\cite {AlessandraSmarter}	
			&		PoW	 &   No
			&		Private	&		Machine learning			\\
			\specialrule{0em}{1pt}{1pt}
			
			Dimobi\textit{ et~al.}\cite {dimobi2020transactive}	
			&		PoW	 &  Yes
			&	Hyperledger	&		-		\\
			\specialrule{0em}{1pt}{1pt}
			
			Van\textit{ et~al.}\cite {van2020integrated}	
			&		-		&  No
			&		Ethereum	&		ADMM			\\
			\specialrule{0em}{1pt}{1pt}
			
			Li\textit{ et~al.}\cite {li2019blockchain}
			&		PoW			&  Yes
			&	 Permissioned  		&		Asymmetric crypto.				\\
			\specialrule{0em}{1pt}{1pt}
			
			Eric\textit{ et~al.}\cite {munsing2017blockchains}
			&		-				&  No
			&		Ethereum		&  ADMM			\\
			\specialrule{0em}{1pt}{1pt}
			
			Li\textit{ et~al.}\cite {li2019decentralized}
			& PoW  &  No
			&		-	 &   Stackelberg game  \\
			\specialrule{0em}{1pt}{1pt}
			
			Li\textit{ et~al.}\cite{li2018blockchain}	
			&		-		& Yes
			&		Ethereum		&  	Bidding algo.				\\ 		
			
			\bottomrule
		\end{tabular}
	\end{threeparttable} \normalsize
\end{table*}

Having reviewed the aforementioned blockchain-based solutions for microgrids,
we obtain the following findings:
\begin{itemize}

\item Microgrids are subsystems of a large power grid, and its users are often communities or islands with relatively close daily connections, which conforms to the characteristics of the consortium blockchain. Therefore, we believe that the integration of blockchain with microgrids is more suitable than with a large power grid.

\item In terms of security, most solutions are based on the trust platform provided by blockchain, and incorporate cryptographic technologies such as blind signature and asymmetric encryption.

\item Power loss is mostly caused by long-distance transmissions and the excessive use of new renewable energy. The regional nature of a microgrid and the distributed structure of a blockchain can solve these problems. Some solutions combine machine learning and ADMM technologies to aggregate and analyze the energy usage data of a microgrid to achieve effective monitoring and forecasting of electricity.

\item Blockchain can provide transparent and unchangeable records for grid users. Its consensus and incentive measures can be exploited to reduce energy consumption. Smart contracts can provide automated management and eliminate trusted third parties. Based on the above characteristics, many schemes combine game theory and the financial characteristics of blockchain to establish auctions and dynamic pricing markets, making energy distribution more reasonable.

\item However, We find that some solutions for security enhancement still have certain flaws and the encryption algorithm used is relatively simple. Advanced techniques such as identity-based encryption, attribute-based encryption, and RSA encryption are rarely adopted in current research. Privacy protection is also mainly guaranteed by signature-based access control, which lacks in-depth research.

\end{itemize}

All in all, the integration of blockchain and microgrid is a complementary combination of two technical fields and there is still much room for further development.

%% file: 8v2g.tex
\section{Blockchain For Electric Vehicles}\label{Vehicle}

\subsection{Current Issues in Electric Vehicles}


The scope of smart grids continues to expand, and electric vehicles are gradually becoming popular, especially in countries with policy supports such as China and Japan. Due to high mobility, electric vehicles can be seen as mobile terminals of the power grid, thereby realizing important functions. This is the so-called V2G technology, which can improve the performance of the power grid in terms of reliability, efficiency, and stability. However, electric vehicles are not perfectly integrated with smart grids, and there exist a series of problems such as energy shortage, security risks, and privacy leak.


As a distributed consensus ledger, blockchain can solve the problems of trust and privacy and demonstrate a great potential in the mobile field. We survey solutions on the combination of electric vehicles and blockchain from two perspectives:  charging/discharging of electric vehicles and energy transmissions of electric vehicles. We elaborate on each category and summarize and compare related technologies in the end.

\subsection{Charging and Discharging of Electric Vehicles}\label{num2.1}


The blockchain technology can help to address the problems of excessive charging load and unstable voltage of electric vehicles. The works \cite {li2019iterative, li2020consortium} propose two models based on the consortium blockchain technology. It is worth mentioning that \cite{li2019iterative} presents a two-layer model, which can realize optimization of charging/discharging transactions. This scheme also considers a mixed integer programming problem, and proposes a new KHA (Krill Herd Algorithm) to solve the problem. 
Moreover, the purpose of the scheme \cite{chen2020blockchain} is to forecast the power consumption in future to efficiently distribute power. It is built on an Ethereum platform, and its core is an auction quotation system and the corresponding power allocation algorithm.


In addition to load issues, the combination of blockchain and V2G can realize cost optimization.
Liu et al.~\cite{liu2018adaptive} optimize cost through the combination of IOEA (Iceberg Order Execution Algorithm) and smart contracts. Similarly, the authors \cite{liu2018blockchain} put forward their own design to optimize cost and price. The technical core of this solution lies in a JD (jump-diffusion) process for energy exchange. The innovation in \cite{ping2020coordinating} is the combination of Ethereum and Lagrange relaxation. This solution can not only maximize charging profit but also prevent malicious collusion attacks.


Blockchain's  consensus mechanism and smart contracts can guarantee security for electric vehicle charging. The design goal of EVChain \cite{firoozjaei2019evchain} is to solve the privacy leakage and hidden security risks in the charging process. It uses the $k$-anonymity technique combined with blockchain, and the electric vehicles in this scheme can share credits based on reputation. Liu et al.~\cite{liu2019proof} propose a novel consensus algorithmic mechanism to resist attacks. This consensus is based on the ONPoB and is specially suitable for V2G environments.
Huang et al.~\cite{HuangLNSC} develop the LNSC system, a perfect security access mechanism without the dependence on third parties. In the aspect of computation costs, this system has a great advantage over traditional ones.
In \cite{zhou2019design}, Zhou et al.~conduct experiments based on an Ethereum platform, which greatly improves security and increases system efficiency. 


As an emerging special application field, smart communities face certain problems in private charging piles and demand response \cite{wang2018contract}.
Like \cite{wang2018contract}, the community often determines the effective range for vehicle charging and discharging. In addition to security, the decentralization of blockchain can facilitate electric pile sharing, thereby improving the utilization efficiency of electric piles~\cite{hou2017resolution}. Su et al.~\cite{su2018secure} investigate the community residents' demand response to electric vehicle charging. They apply the PBFT consensus and smart contracts to utilize renewable energy, making the charging process more rational.
   
\subsection{Energy Transmissions Based on Electric Vehicles}\label{num2.2}

\begin{figure}[htbp]
	\includegraphics[width=.95\linewidth]{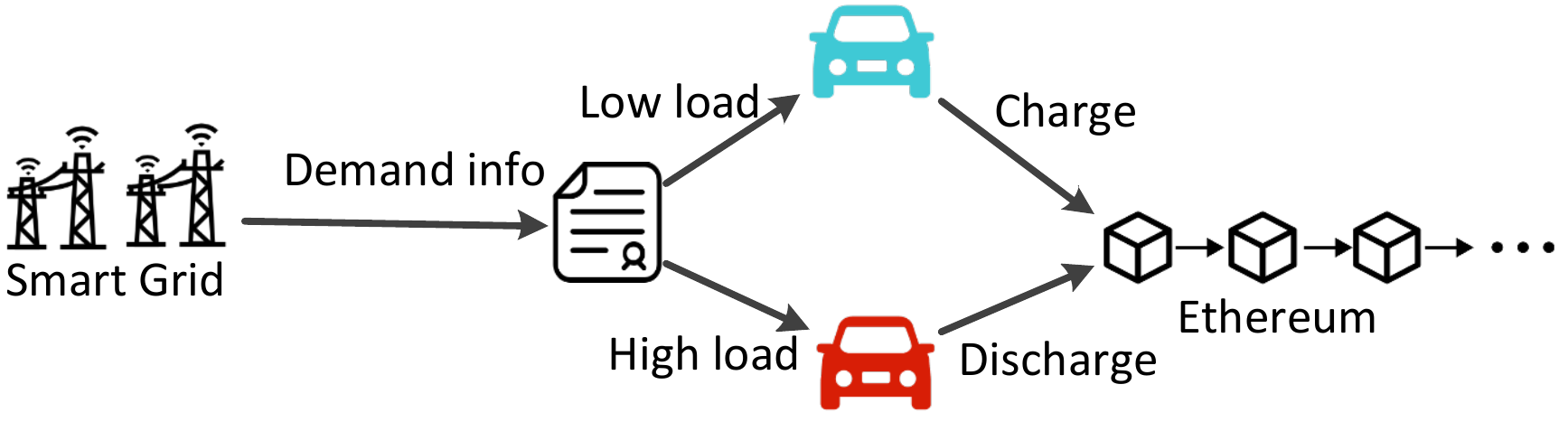}
	\caption{The model in \cite{GuoDistributed} involves three components: vehicle, blockchain network and power grid.}
	\label{fig:car}
\end{figure}

A smart grid can provide transit points for energy transactions through electric vehicles, which greatly improves the distance and flexibility of energy transmissions \cite{GuoDistributed} (as illustrated in Figure \ref{fig:car}). It can provide energy for edge base stations or nodes to solve the problem of insufficient energy at edges when necessary. Blockchain can provide trust, security, and privacy for the energy transmissions of V2G \cite{Kim2018Blockchainfor}.

Electric vehicles, as movable energy transmission nodes, combined with blockchain technology can reduce the cost of long-distance transmissions.
Ref. \cite {kang2017enabling} proposes a power trading model based on a consortium blockchain. The model mainly implements local power transactions between PHEVs in a smart grid. In addition, it employs an auction mechanism to solve the transaction pricing problem. Hassija et al.~\cite{hassija2020blockchain} apply DAG to improve the efficiency of system operation, and meanwhile combine it with game theory to achieve rational pricing between vehicles and the power grid. Liu et al.~\cite{liu2019electric} make use of the transparency and immutability properties of Ethereum to implement a price auction mechanism. Smart contracts in this scheme are used to realize dynamic pricing. Ref. \cite {xue2020design} employs game theory to balance the needs and interests of energy trading platforms.
The works \cite{zhou2019secure, zhou2018blockchain} combine blockchain, edge computing, and game theory to achieve fair pricing. On this basis, an ICC (iterative convex-concave) algorithm is designed to maximize the welfare of the entire system.


The combination of electric vehicles and blockchain can inherit their respective advantages and achieve demand response during energy transmissions. 
The work \cite{islam2020blockchain} discusses the situation where renewable energy resources cannot meet vehicle needs at peak times; and the authors put forward their own opinions at the end of the paper. In \cite{liu2019peer}, the main contribution is to propose a new PoB consensus algorithm to stimulate demand response. The new consensus algorithm can be implemented on Ethereum and combined with smart contracts to eliminate third parties.


Moreover, blockchain can provide security for V2G. 
Sheikh et al.~\cite{sheikh2019secured} mainly address the false data attack problem. The proposed scheme improves the PBFT consensus and confirms the feasibility of the scheme based on the standard IEEE33 bus system. Wang et al.~\cite{wang2019bsis} propose an incentive mechanism called BSIS for smart cities. This mechanism combines a credit-based consensus and the cryptocurrency technology, which can effectively prevent external attacks.

In addition to resist external attacks, the security risks brought by third parties cannot be ignored.
Bansal et al.~\cite{BansalSmartChain} point out that the blockchain technology can provide privacy protection, contribute an effective consensus mechanism and eliminate dependence on trusted third parties. As the computing resources of electric vehicles are limited, the authors design SmartChain, a smart and scalable ledger framework inspired by blockchain.
Sun et al.~\cite{sun2020blockchain} employ a consortium blockchain to eliminate third parties. Compared with other schemes, this approach improves the Byzantine algorithm based on DPoS consensus. The authors also design a new auction mechanism to optimize prices.


Both \cite{jindal2019survivor} and \cite{chaudhary2019best} propose security solutions based on SDN networks. The solution reported by Jindal et al.~\cite{jindal2019survivor} is named SURVIVOR, which randomly selects validators to perform security verification on all transactions. Chaudhary et al.~\cite{chaudhary2019best} present a scheme named BEST. The security of this scheme is also verified based on special nodes. This scheme is mainly used in the field of autonomous driving.

\subsection{Summary}

We summarize the design goals, key technologies and other blockchain-related attributes of the solutions outlined in this section and report the results in Table \ref{vehicle22}.

\begin{table*}[htbp]
	\begin{threeparttable} \footnotesize
		\caption{Blockchain For Electric Vehicle}
		\vspace{-0.3cm}
		\label{vehicle22}
		\begin{tabular}{c c c c c c c c c}
			\toprule
			Ref.   & Design goal   &   Consensus      &   Blockchain & Contract       \\
			
			\specialrule{0em}{1pt}{1pt}
			\midrule
			
			Su\textit{ et~al.}\cite {su2018secure} & Security and privacy
			&  RDBFT
			&		Permissioned 	&	 Used 				\\
			\specialrule{0em}{1pt}{1pt}
			
			Zhou\textit{ et~al.}\cite {zhou2019design}		& Security and privacy
			&		Hybrid		
			&   Ethereum		&   	Used  		 \\
			\specialrule{0em}{1pt}{1pt}
			
			Wang\textit{ et~al.}\cite {wang2018contract}   & Security and privacy  
			&  RDBFT		
			&		Permissioned 	&	 Used 					\\
			
			Liu\textit{ et~al.}\cite {liu2019proof} 		&	Security and privacy  
			&		PoB	 
			&		Ethereum		&			Used				\\
			\specialrule{0em}{1pt}{1pt}

			BEST\cite {chaudhary2019best} 	&	Security and privacy
			& PoW 
			&		Consortium 	& 		Unused  \\
			\specialrule{0em}{1pt}{1pt}
			
			Li\textit{ et~al.}\cite {li2019iterative}	&	Load optimization	
			&		 Kafka				
			&		Hyperledger			&			Unused			\\
			\specialrule{0em}{1pt}{1pt}
			
			Li\textit{ et~al.}\cite {li2020consortium}	& Load optimization		
			&		 Kafka				
			&		Hyperledger		&			Unused			\\
			\specialrule{0em}{1pt}{1pt}
			
			Kang\textit{ et~al.}\cite {kang2017enabling}  & Cost optimization 
			&  PoW 
			&     Consortium    &      Unused         \\
			\specialrule{0em}{1pt}{1pt}
			
			Liu\textit{ et~al.}\cite {liu2019electric}	&	Cost optimization 
			&		PoA			
			&		Ethereum	&			Used						\\
			\specialrule{0em}{1pt}{1pt}
			
			SmartChain\cite {BansalSmartChain}	& Third party	
			&   Smartchain		&   Smarrchain 	&   	Unused 	  \\
			\specialrule{0em}{1pt}{1pt}
			
			Sun\textit{ et~al.}\cite {sun2020blockchain} 	&	Third party
			& DPoSP 
			&	Consortium  	& 		Unused 	  \\
			\specialrule{0em}{1pt}{1pt}
			
			Zhou\textit{ et~al.}\cite {zhou2019secure}  	&	Edge computing	
			&		PoW			
			&	 Consortium 	&			Used				\\
			\specialrule{0em}{1pt}{1pt}
			
			Zhou\textit{ et~al.}\cite{zhou2018blockchain} &		Edge computing		
			&		PoW			
			&	 Consortium 	&			Used					\\
			\specialrule{0em}{1pt}{1pt}
			
			Sheikh\textit{ et~al.}\cite {sheikh2019secured} 	&	Anti-attack
			& Byzantine 
			&		Ethereum  	& 		Used 	\\
			\specialrule{0em}{1pt}{1pt}
			
			BSIS\cite {wang2019bsis} 	&	Anti-attack
			& PoR 
			&		Permissioned  	& 		Unused 	\\
			\specialrule{0em}{1pt}{1pt}
			
			Liu\textit{ et~al.}\cite {liu2019peer} 	&	Demand response 
			& 	PoB			
			& Ethereum	&   Used         \\
			
			\bottomrule
		\end{tabular}
	\end{threeparttable} \normalsize
\end{table*}

Having reviewed the aforementioned blockchain-based solutions for V2Gs,
we obtain the following findings:
\begin{itemize}

\item We conclude that using blockchain-based solutions to augment electric vehicles can provide safe charging/discharging and power transmission services. In the process of charging/discharging, the use of blockchain can make the private electric piles public and improve the resource utilization of the entire system. 
In the process of energy transmission, vehicles can transmit electrical energy to remote areas during peak power periods, ensuring the stability of the entire system. Blockchain can provide vehicles with a transparent energy management platform.

\item More specifically, the combination of blockchain and V2G is mainly to deal with security risks, demand response, and process optimization. Security risks are mainly caused by external attacks, privacy leaks and third parties. Many solutions based on blockchain combine ECDSA, asymmetric encryption, and other cryptographic algorithms to greatly improve the security level of the system. Process optimization can be considered from the aspects of cost optimization, load optimization, service optimization, etc. Blockchain's own trust and financial attributes can attract a number of vehicles while reducing price fluctuations. Many schemes combine game theory and various improved algorithms on this basis to further realize optimization.

\item With the rapid increase in electric vehicles and the high complexity of traffic management, traditional V2G management consumes a lot of manpower and costs. The use of smart contracts can replace labor and realize automated operations of the system.

\item The diversity of blockchain platforms can provide application scenarios for different solutions. Due to the real-time requirements of V2G and the limitation of vehicle computing power, many solutions have abandoned the traditional PoW consensus and proposed various consensus algorithms such as Kafka, DPoSP and PoB, which effectively improve the efficiency of the system. 

\item However, the problem of blockchain transmission delay in long-distance networks still exists. Whether the vision of blockchain joining V2G can be supported by most people also requires deeper thinking and further research.

\end{itemize}

%% file: 9resource.tex
\section{Current Projects and Initiatives}\label{resource}

For completeness of this survey, we summarize current projects and initiatives that apply blockchain to smart grids. 

\textbf{Carbon Asset Development Platform (C.P) \cite{EnergyChainTechnology}}. This platform \cite{EnergyChainTechnology} is proposed by Energy Chain Technology Corporation and IBM, which is mainly used in the carbon market to achieve energy saving, emission reduction and green energy. The blockchain is intended to realize the traceability and transparent management of data, so as to further realize the validity and authenticity of information.

\textbf{Russian Carbon Market (R.C.M) \cite{IPCI}}. In response to the call for green energy, Russian company DAO IPCI employs blockchain technology to facilitate the business of local carbon industry companies. This platform is safe and open, and all nodes can monitor the data on the blockchain. In the implementation process, Hinprom takes the lead in experimental studies. 

\textbf{Israel Microgrid (I.M) \cite{Israelcleanenergy}}. The Israel microgrid makes use of blockchain to realize the issuance of green certificates. The platform proposed by Greeneum can guarantee the correctness of energy production and then issue corresponding green certificates. On this basis, the project  combines artificial intelligence to achieve electricity prediction and balance, making P2P transactions more reasonable.


\textbf{Energy Web Foundation (E.W.F) \cite{Energyweb}}. Energy Web Foundation takes advantage of the underlying structure of Ethereum to integrate more than 100 power companies. Recently, it  proposes EW-DOS to create a decentralized operating system for accelerating the construction of low-carbon power systems.

\textbf{Tanzania X-Solar \cite{xsolarsystem}}. The X-Solar project is provided by M-PAYG with the corresponding technical support. The purpose of X-Solar is to solve the power shortage problem in developing countries of Africa such as Tanzania. It can provide solar power monitoring, as well as transaction and payment functions based on the blockchain.

\textbf{Iberdrola Energy Tracking (I.E.T) \cite{Iberdrola}}. This project is established by Iberdrola and Kutxa investment, and has been successfully tested. It employs the blockchain technology to realize the management and monitoring of energy trading processes. In addition, blockchain is used to provide trusted energy certificates.

\textbf{Poseidon Foundation (P.F) \cite{Poseidon}}. The Poseidon foundation focuses on reducing carbon emissions. It cooperates with Ben \& Jerry's to combine carbon trading with ice cream, effectively reducing carbon emissions. Stellar blockchain can improve the transparency of transactions. At present, it is actively working with the LCC (City Council of Liverpool) to build a carbon credit mechanism for reducing carbon emissions and protecting the environment of London.

\textbf{TenneT and IBM Pilot (T.I.P) \cite{TenneT}}. In Germany, TenneT, IBM and Sonnen have cooperated to realize distributed storage of electricity. This effectively solves the high-cost management caused by excess energy supply. This project utilizes Hyperledger to record and trace the corresponding data information.
An application of this project in electric vehicle charging in the Netherlands has also achieved certain results.

\textbf{Brooklyn Microgrid (B.M)\cite{Brooklynmicrogrid}}. In Brooklyn, the local power grid company (LO3) proposes to combine  Ethereum with the smart grid and implement it locally. It allows residents to use solar energy to generate electricity and trade to obtain token rewards.

\textbf{IDEO CoLab Solar Panel (I.C.S.P) \cite{IDEOCoLab}}. This is a joint project between IDEO CoLab and Filament. They aim to combine solar panels with the Internet of Things, and the bottom layer uses blockchain and smart contracts for monitoring, energy production, and automatic transactions.

\textbf{Global Grid (G.G) \cite{globalgrid}}. Global Grid cooperates with a local Mexican developer to build a solar blockchain platform and a large photovoltaic grid. It is understood that this project also intends to expand to other renewable resource areas such as geothermal.

\textbf{Electron \cite{Electron}}. Electron conducts community experiments in London based on Ethereum and IPFS. Residents can trade electricity with their neighbors through Electron.

\textbf{Scanergy \cite{SCANERGY}}. This project is launched by the European Union, which aims to achieve green energy trading. Scanergy is based on Ethereum and smart contracts, and uses NRGcoin as the digital currency. It also makes use of the cloud market to derive a transparent energy auction platform.

\textbf{Fremantle Project (F.P) \cite{Fremantlepowerledger}}. The Fremantle Project aims to optimize the use of distributed energy and water resources in Australia. Blockchain is employed to collect digital information of various energy sources to ensure its integrity. 

\textbf{De La Salle University Microgrid (D.L.S.U.M) \cite{Philippinesmiero}}. This is a microgrid system constructed inside the De La Salle University. This system uses photovoltaic power generation, combined with the Qtum blockchain system for P2P transactions. The actual effect shows that the system can effectively realize the optimization of cost, environment, and safety.

\textbf{PONTON Enerchain \cite{Ponton}}. PONTON is the underlying blockchain platform of Enerchain, which supports the electricity trading and transmission of more than 30 energy companies in Europe.


\textbf{Solar Bankers (S.B) \cite{Solarbankers}}. Solar Bankers is a distributed solar trading platform proposed by Singapore. In Solar Bankers, blockchain is adopted to guarantee the security and transparency of solar energy production and transaction processes. Users can obtain the digital currency SunCoin by trading energy.

\textbf{Rotterdam Heat Network (R.H.N) \cite{Eneco}}. CGI \& Eneco designs a heat energy trading platform based on Tendermint and conducts a field test in Rotterdam.

\textbf{Austria Grid Project (A.G.P) \cite{GridSingularity}}. Grid Singularity is an Austrian energy trading platform. It has conducted experiments on the smart grid locally. It is implementing an open and secure distributed smart grid system.

\textbf{Interbit \cite{BTLgroup}}. Interbit is a blockchain system created by BTL Group. It has conducted field experiments in the United Kingdom, Canada, and other places, and achieves good results in energy trading.

\textbf{TEPCO \& Blockchain (T.B) \cite{electronTEPCO}}. TEPCO is a Tokyo-based power company that has worked with multiple companies such as Grid+ to create a new type of distributed smart grid. It creates a P2P solar trading system together with Conjoule. 
Later, TEPCO joins the EWF (Energy Web Foundation) to accelerate the development of the global energy blockchain and collaborates with Electron on the decentralized transformation of smart grids.

\textbf{Hoog Dalem LEF (H.D.L) \cite{energy21}}. It is a pilot project jointly cooperated by Stedin, Energy21, ABB, and iLeco. In this project, residents conduct P2P transactions and settle through blockchain.

\textbf{SunContract \cite{SunContract}}. SunContract is a P2P energy trading platform supported by the European Union. The test in Slovenia shows that it can effectively reduce energy cost.


\textbf{Share \& Charge (S.C) \cite{sharecharge}}. The main application area of the Share \& Charge platform is V2G. It makes the free private charging piles public. Each mobile app acts as an OCN node. Blockchain makes transactions transparent and open, and enhances the credibility of the system.

\textbf{Oxygen Initiative (O.I) \cite{oxygeninitiative}}. Similar to Share \& Charge, Oxygen Initiative is also used in the charging field of vehicles. The predecessor of the Oxygen Initiative is the vehicle e-wallet project of RWE and ZF company. It is based on the Ethereum platform, which can provide secure payment services for vehicle expenses, and has been simulated in the US electricity market.

\textbf{JuiceNet \cite{JuiceNet}}. JuiceNet is a marketing platform created by eMotorWerks. The platform realizes the charging function of P2P based on blockchain. Electric vehicles can be charged via other people's private charging piles according to demand, and pay the corresponding fees. This not only provides charging convenience for vehicles, but also encourages more residents to purchase charging piles to earn fees.

We summarize the projects and initiatives mentioned above in Table~\ref{Resource}.

\begin{table*}[htbp]
	\begin{threeparttable} \footnotesize
		\caption{Comparison: Practical Applications and Public Experiments}
		\vspace{-0.3cm}
		\label{Resource}
		
		\begin{tabular}{c c c c c c}
			\toprule
			Design Goal       & 	Ref.    & Organization  &   	Energy type     &   Blockchain      \\
			\specialrule{0em}{1pt}{1pt}
			\midrule
					& 	C. P \cite{EnergyChainTechnology}	&  -  	&	ECTC/ IBM				&	 Carbon  	&     -	\\
		\specialrule{0em}{1pt}{1pt}
		
		
		& 	R.C.M \cite{IPCI} 	&		Russian		&	DAO IPCI			&				Carbon			&	-	\\
		\specialrule{0em}{1pt}{1pt}
		
		& I.M \cite{Israelcleanenergy} 	&	Israel	&	Greeneum		&			 Green energy				&		Ethereum	\\
		\specialrule{0em}{1pt}{1pt}
		
		
		& 	E.W.F \cite{Energyweb} 	& Switzerland	&	Energy Web Foundation							&  Electricity    &   Ethereum 		\\
		
		Management		& 	X-Solar \cite{xsolarsystem}	& Tanzania	&	M-PAYG 					& Solar    &   -		\\
		\specialrule{0em}{1pt}{1pt}
		
		& 	I.E.T \cite{Iberdrola}  	& Spain &	Iberdrola/ Kutxa			&  Renewable    &   -		\\
		\specialrule{0em}{1pt}{1pt}
		
		& 	P.F  \cite{Poseidon}	& Liverpool  &	LCC			& Carbon    &   Stellar		\\
		\specialrule{0em}{1pt}{1pt}
		
		& 	T.I.P  \cite{TenneT}	&  Germany 	&	TenneT/ IBM/ Sonnen		&	Electricity    &   Hyperledger \\
		\specialrule{0em}{1pt}{1pt}
		
		\midrule
		
		& B.M \cite{Brooklynmicrogrid} &  Brooklyn  & LO3  &     Solar            &   Ethereum   \\
		
		\specialrule{0em}{1pt}{1pt}
		
		& I.C.S.P  \cite{IDEOCoLab} &    -  	 &  IDEO CoLab/ Filament   		   &    Solar     &     -  \\
		\specialrule{0em}{1pt}{1pt}
		
		&	G.G \cite{globalgrid} &		Mexico	&		SAPI de CV							&	Solar 					&		-	 \\
		\specialrule{0em}{1pt}{1pt}
		
		& Electron \cite{Electron}	&  London  &	Electron			  		&   Electricity  						&		Ethereum 	\\
		\specialrule{0em}{1pt}{1pt}
		
		&	Scanergy \cite{SCANERGY} &		Europe	&		-				&		Green energy				&		Ethereum	\\
		\specialrule{0em}{1pt}{1pt}
		
		
		
		
		
		
		&	F.P \cite{Fremantlepowerledger} 	& Australia &	Australian government				  		&	Distributed			     &   -  \\
		\specialrule{0em}{1pt}{1pt}
		& D.L.S.U.M \cite{Philippinesmiero}	&  Philippines  &	Energo Labs				  		&	Solar			    &   Qtum  \\
		\specialrule{0em}{1pt}{1pt}
		
		Trading &  Enerchain \cite{Ponton}  &     Germany  & PONTON      &   Electricity       &      Tendermint     \\
		\specialrule{0em}{1pt}{1pt}
		
		
		& S.B \cite{Solarbankers}	&	Singapore  &  Solar Bankers  			 		&	Solar 						&  Skyledger 	\\
		\specialrule{0em}{1pt}{1pt}
		
		&	R.H.N \cite{Eneco}	& Rotterdam	&	CGI \& Eneco						&  Heat     &   Tendermint 		\\
		\specialrule{0em}{1pt}{1pt}
		
		& A.G.P \cite{GridSingularity}	& 	Austria 	&	Grid Singularity			&  Electricity    &   - 		\\
		\specialrule{0em}{1pt}{1pt}
		
		&	Interbit \cite{BTLgroup}	& UK	&	BTL Group					&  Electricity    &   Interbit		\\
		\specialrule{0em}{1pt}{1pt}
		
		&	T.B  \cite{electronTEPCO}	& Tokyo  &	Conjoule/ Electron		 &   Electricity   &   Ethereum		\\
		\specialrule{0em}{1pt}{1pt}
		
		& H.D.L \cite{energy21}	& Netherlands  &	Energy21/ Stedin			&   Electricity   &   Quasar		\\
		\specialrule{0em}{1pt}{1pt}
		
		&	SunContract \cite{SunContract}	& Slovenia  &	SunContract			&   Electricity   &   Ethereum		\\
		
		
		\midrule
		
		& S.C \cite{sharecharge} &  	 Germany 		&  Innogy/ Slock.it  &    		Electricity				&		Ethereum 	\\
		\specialrule{0em}{1pt}{1pt}
		
		V2G		& O.I \cite{oxygeninitiative} &		US	&	Oxygen Initiative		&			Electricity 				&		Ethereum	\\
		\specialrule{0em}{1pt}{1pt}
		
		&	JuiceNet \cite{JuiceNet} 	& North America	&	eMotorWerks				& Electricity    &   -		\\
		\specialrule{0em}{1pt}{1pt}
		
		
		
			\bottomrule
		\end{tabular}
	\end{threeparttable} \normalsize
\end{table*}

%% file: 10summary.tex
\section{ Open Issues}

Although blockchain has been applied in many fields, it still faces huge challenges requiring effective solutions when adopted by smart grids. In this section, we highlight some of these challenges and point out future directions to stimulate further research.


\noindent{\bf Reliability and safety}. Since a power grid is a critical infrastructure,
its reliability and safety are of paramount importance for all energy providers and consumers.
The smart grid has used digital information and control mechanisms such as load adjustment/balancing, demand-side management, and self-healing technologies to achieve reliability and safety.
Blockchain can be used to enhance reliability and safety of a smart grid,
but this demands a deep integration of blockchain
with the smart grid. Such an integration requires interdisciplinary collaborations among
experts in computer science, information security, and electric systems.

The inherent fault tolerance of blockchain can help to enhance reliability of smart grids.
Blockchain can be applied to components of a smart grid
like the intelligent control system and the operational system to make them more robust against
faults or attacks. Meanwhile, because of the immutability of blockchain, it can be
an ideal media to share critical control information for grid management.

\noindent{\bf Performance}. A smart grid is featured by real-time control, monitoring,
and information exchange, and it has strict performance requirements on
the communication latency. Unfortunately, blockchain is not designed
for real-time applications, since its consensus latency may be too long for control,
optimization or trading in smart grids. Even fast consensus algorithms such as DPoS
still require seconds to reach a consensus. How to design a blockchain system
that can process data in real time is both interesting and challenging.

It is also important to improve computation and communication efficiency of blockchain
to support real-time operations in smart grids. The blockchain validators/miners
need to synchronize and verify each transaction and block, which may introduce long delays for smart grids.
Thus efficient algorithms should be specifically designed for real-time operations in smart grids.

\noindent{\bf Scalability}. As a smart grid consists of many actuators, measurement units, sensors
and smart meters that scatter over a wide area, the blockchain system must be highly scalable to efficiently and speedily manage them. This calls for a scalable blockchain architecture design to support
large-scale smart grids. Such a consideration has not received enough attention from the research community.

Fortunately, many potential solutions have been proposed recently to improve blockchain
throughput. Blockchain sharding has been a very effective mechanism
to improve scalability of a blockchain system. The multi-chain mechanism divides
the entire system into independent components, with each running its own blockchain.
Sidechain and the corresponding pegged mechanism can also contribute to improve scalability.
For both multi-chain and sidechain, appropriate cross-chain protocols should be applied to ensure security for transactions.
The off-chain or layer-2 solutions such as Lightning Network may be used
to shorten the confirmation delay of transactions in blockchain.

\noindent{\bf Security issues}.
Blockchain can help to establish trust and it is deemed trustworthy, but inappropriate use of
the blockchain technology may lead to severe consequences, especially for smart grids.

Like most security-critical information systems, the most important security problem in a smart grid is how to securely manage cryptographic keys. However, many existent blockchain systems
do not adopt effective key management strategies, and they just allow users to generate random keys
without certification. This can result in severe security problems, e.g. impersonation attacks and sybil attacks.
Entity and data authentication are basic requirements for security of smart grids, and this is only possible when
effective key management mechanisms are applied to blockchain.

A number of security threats should be carefully studied in designing a blockchain system for smart grids.
Illegal access to critical data of smart grids should be effectively prevented with access control mechanisms,
which are important for data sharing in smart grids.
Hardware security should be considered in case of device compromise in smart grids.
For example, devices like smart meters are easy to be compromised and tamper-proof hardware may be used to guarantee
data security. For electricity trading, fairness should be guaranteed to thwart malicious consumers or providers.

\noindent{\bf Privacy issues}.
Privacy issues are even more challenging than security for both blockchain and smart grids.
Blockchain is well-known for its transparency and public verifiability, which means {\em every}
validator/miner can access the content of all transactions. Different types of privacy need to be
protected, including identity privacy, location privacy, amount privacy, relationship privacy,
and data privacy.

From the perspective of privacy protection,
the single point of failure in a centralized system is turned into {\em multiple} single points
of failures in a blockchain system. An attacker only needs to compromise one validator/miner to
obtain sensitive information of the smart grid. Therefore, the privacy problem is both challenging
and critical when applying blockchain to smart grids.
In addition, traffic analysis is also a potential threat to secure operations of smart grids.

Simple encryption cannot solve privacy issues as validators/miners still need to verify the blockchain data. Advanced cryptographic techniques can be applied to solve privacy issues. Ring signature,
zero knowledge proof, zk-SNARKs, homomorphic encryption have been proposed recently for privacy preservation in blockchains.
However, these techniques are normally inefficient in computation and/or communication, so
they must be carefully adapted to meet the requirements of smart grids.

\noindent{\bf Compatibility for low-end devices}. The actuators, sensors, smart meters and measurement units
in a smart grid are usually resource-constrained, and it is difficult for them to carry out
certain public key cryptographic computations. The security mechanisms should accommodate
these low-end devices, and the computation burden on these devices should be minimized without
losing security and privacy. For example, low-end devices should only carry out
symmetric encryption and hashing operations, and should avoid frequent signature generation/verification.

%% file: 11conclusion.tex
\section{Conclusion}
In this survey, we have reviewed the latest research progress on applying the blockchain technology to smart grids.
We first summarize characteristics of blockchain systems, the smart contract paradigm, and
the advantages of the blockchain technology. Then we categorize the relevant research works into
5 groups: intelligent energy management, energy trading, security and privacy, microgrid management, and electric vehicles.
For each category, we briefly discuss issues with traditional solutions without using blockchain.
Then we analyze and compare how blockchain is utilized to tackle the issues. A table is provided to give a clear comparison among these solutions, followed by a brief summary for each category.
For completeness of this survey we also list and summarize related projects and initiatives.
Finally, we elaborate on open problems and research challenges that have not been
sufficiently investigated by current research.

Apparently, the blockchain technology has many desirable properties that can be utilized to
construct a better smart grid. To this end, collaborations among professionals from different areas
are very important. We expect that this survey can provide principle reference and direction guidance for the future development of blockchain-enabled smart grids.